# Single star progenitors of long gamma-ray bursts I:

## Model grids and redshift dependent GRB rate

S.-C. Yoon[1], N. Langer[2], and C. Norman[2,3,4]

[1] Astronomical Institute "Anton Pannekoek", University of Amsterdam, Kruislaan 403, 1098 SJ, Amsterdam, The Netherlands
e-mail: scyoon@science.uva.nl
[2] Astronomical Institute, Utrecht University, Princetonplein 5, NL-3584 CC, Utrecht, The Netherlands
e-mail: n.langer@astro.uu.nl
[3] The Johns Hopkins University, Homewood Campus, Baltimore, MD 21218, USA
[4] Space Telescope Science Institute, 3700 San Martine Drive, Baltimore, MD 21218, USA
e-mail: norman@stsci.edu

**Abstract.** We present grids of massive star evolution models at four different metallicities ($Z = 0.004, 0.002, 0.001, 0.00001$). The effects of rotation on the stellar structure and the transport of angular momentum and chemical elements through the Spruit-Tayler dynamo and rotationally induced instabilities are considered. After discussing uncertainties involved with the adopted physics, we elaborate the final fate of massive stars as a function of initial mass and spin rate, at each considered metallicity. In particular, we investigate for which initial conditions long gamma-ray bursts (GRBs) are expected to be produced in the frame of the collapsar model. Then, using an empirical spin distribution of young massive metal-poor stars and a specified metallicity-dependent history of star-formation, we compute the expected GRB rate as function of metallicity and redshift based on our stellar evolution models. The GRB production in our models is limited to metallicities of $Z \lesssim 0.004$, with the consequence that about 50 % of all GRBs are predicted to be found at redshifts above $z = 4$, with most supernovae occurring at redshifts below $z \simeq 2.2$. The average GRB/SN ratio predicted by our model is about 1/200 globally, and 1/1250 at low redshift. Future strategies for testing the considered GRB progenitor scenario are briefly discussed.

## 1. Introduction

Rotation is known to affect the evolution of massive stars significantly, through the resulting centrifugal force and through rotationally induced chemical mixing (Maeder & Meynet 2000; Heger et al. 2000). In particular, very efficient chemical mixing may persist in massive stars when the mixing time scale decreases below the thermonuclear time scale in very rapid rotators. In this situation, a strong chemical gradient can never be established. As a result, the star remains quasi-chemically homogeneous and evolves bluewards (Maeder 1987; Langer 1992), contrary to slower rotators which develop the classical core-envelope structure and evolve redwards. This chemically-homogeneous evolution scenario (hereafter, CHES) has been invoked to understand nitrogen-rich (Howarth & Smith 2001; Bouret et al. 2003; Walborn et al. 2004), and helium-rich (Mokiem et al. 2006) massive main sequence stars in the Magellanic Clouds.

Only recently, the CHES is recognized as a promising path towards collapsars in connection with long gamma-ray bursts (GRBs). The collapsar scenario requires massive helium stars with rapidly spinning cores ($j \gtrsim 10^{16}$ cm$^2$ s$^{-1}$; Woosley 1993). Stellar models computed with magnetic torques generally fail to retain such high core angular momenta (Heger, Woosley & Spruit 2005; Petrovic et al. 2005), while they can consistently explain the spin rates of young neutron stars (Heger, Woosley & Spruit 2005; cf. Ott et al. 2006) and white dwarfs (Suijs et al. 2006), as well as the internal rotational profile of the Sun (Eggenberger, Maeder & Meynet 2005). However, Yoon & Langer (2005; YL05) and Woosley & Heger (2006; WH06) showed that at low metallicity, quasi-chemically-homogeneous evolution of rapidly rotating massive stars can lead to the formation of rapidly rotating massive helium stars which satisfy all the requirements of the collapsar scenario even if the effect of magnetic torques is included. This is possible since mixing the hydrogen-rich envelope into the core rather than losing it to a wind avoids the angular momentum loss associated with mass loss (Langer1998), and avoiding a core-envelope structure prevents the magnetic core-envelope coupling and the corresponding core spin-down.

*Send offprint requests to*: S.-C. Yoon

The CHES for GRB progenitors predicts that GRBs should occur preferentially in metal poor environments (YL05; WH06), which seems to be confirmed by recent observations. Most GRB host galaxies appear to have sub-solar metallicity, even down to $Z \simeq Z_{\odot}/100$ (e.g. Fynbo et al. 2003; Conselice et al. 2005; Gorosabel et al. 2005; Chen et al. 2005; Starling et al. 2005; Fruchter et al. 2006; Fynbo et al. 2006; Stanek et al. 2006; Mirabal et al. 2006; Wiersema et al. 2006). The use of GRBs as probes of star formation at high redshift thus demands a quantitative understanding of their low-metallicity bias, which to provide in the frame of the CHES is a major goal of this paper.

To this purpose, we present grids of stellar evolution models at $Z$ =0.004, 0.002, 0.001, and 0.00001, for rotating magnetized stars in the initial mass range $12 \leq M_{\rm init}/{\rm M}_{\odot} \leq 60$, and with initial equatorial rotation velocities between zero and 80% of the Keplerian value ($0 \leq {\rm v}_{\rm init}/{\rm v}_{\rm K} \leq 0.8$). Our numerical methods are described in the next section (Sect. 2), and physics uncertainties are critically discussed in Sect. 3. In Sect. 4, we present the computed stellar evolution models, and discuss the final fate of massive stars as function of initial mass, rotational velocity and metallicity. The GRB rate as a function of metallicity and redshift following from our models is addressed in Sect. 5. After a discussion of our results in Sect. 6 we summarize our conclusions in Sect. 7.

## 2. Numerical methods and physical assumptions

We use the same hydrodynamic stellar evolution code as in YL05, which includes the effect of rotation on the stellar structure, transport of angular momentum and chemical species via magnetic torques and rotationally induced hydrodynamic instabilities, with several improvements. Uncertainties introduced by the processes discussed in this section are elaborated in Sect. 3.

### 2.1. Mixing

We adopt a larger value for the semi-convective mixing parameter ($\alpha_{\rm SEM} = 1.00$) than in YL05 (where $\alpha_{\rm SEM} = 0.04$). This choice facilitates comparison of our results with models of other groups as it yields stellar core sizes comparable to those of WH06, and of the Geneva group who adopts the Schwarzschild criterion (see discussion in Langer et al. 1985; Langer 1991). Uncertainties involved with the value of $\alpha_{\rm SEM}$ are discussed in Sect. 3.4.

YL05 followed Heger, Langer & Woosley (2000) for the calibration of the efficiency of the rotationally induced chemical mixing, which was fit to observed surface nitrogen and helium abundances of Galactic O stars. However, as the current version of the code incorporates new physics such as the effects of magnetic torques (Petrovic et al. 2005) and the use of a larger semi-convection efficiency parameter, we re-calibrated the mixing efficiency accordingly. Specifically, a larger value for the correction factor for the effect of mean molecular weight gradients on rotational mixing ($f_{\mu} = 0.1$) is now employed, compared to the previously used value ($f_{\mu} = 0.05$; Heger et al. 2000).

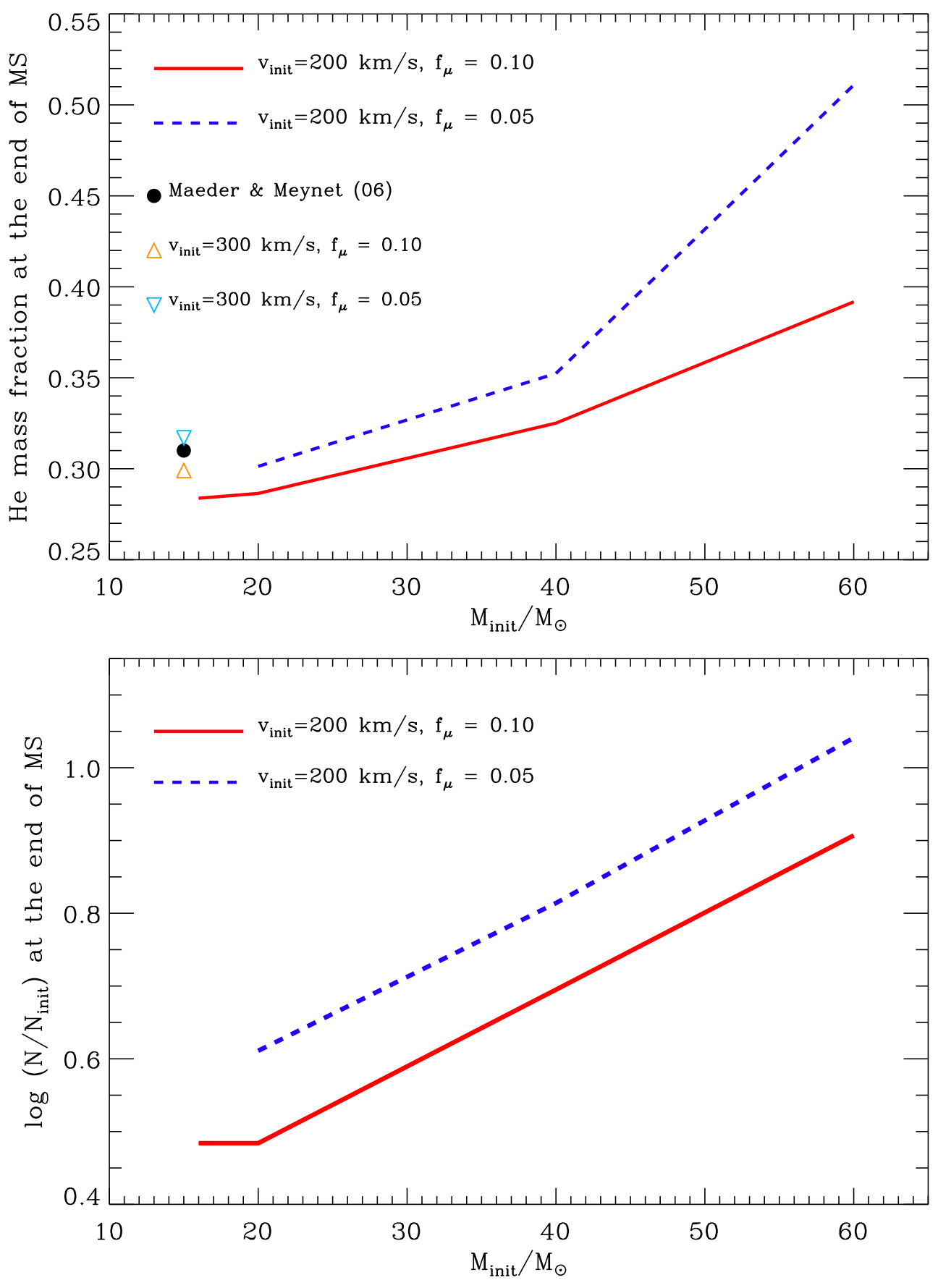

**Fig. 1.** *Upper panel*: Surface helium mass fraction in stellar models (Z=0.02) at core hydrogen exhaustion, as function of the initial stellar mass. *Lower panel*: Logarithm of the surface nitrogen abundance divided by its initial value, at the end of core hydrogen burning, for the same models as shown in the upper part.

As shown in Fig. 1, at core hydrogen exhaustion in models with an initial equatorial rotation velocity of ${\rm v}_{\rm init} = 200\,{\rm km\,s^{-1}}$, the use of $f_{\mu} = 0.1$ gives surface abundances of nitrogen and helium comparable to those in the corresponding models by Heger et al. with their fiducial value of $f_{\mu} = 0.05$ (i.e., $Y_{\rm s} = 0.29 \sim 0.40$ and $\log N/N_{\rm init} = 0.5 \sim 0.9$ for $M_{\rm init} = 20 \sim 60\ {\rm M}_{\odot}$). In the same figure, we also compare our result with a magnetic model of Maeder & Meynet (2005): for $f_{\mu} = 0.1$, the surface helium enrichment at core hydrogen exhaustion in a sequence with $M_{\rm init} = 15\ {\rm M}_{\odot}$ and ${\rm v}_{\rm init} = 300\ {\rm km\,s^{-1}}$ is just slightly less in our model ($Y_{\rm s} = 0.299$) compared to Maeder & Meynet ($Y_{\rm s} = 0.310$).

### 2.2. Mass loss

We follow Kudritzki et al. (1989) for computing the stellar wind mass loss of hot, hydrogen rich stars, with a metallicity dependence of $\dot{M} \propto Z^{0.69}$, as suggested by Vink et al. (2001). Here $Z$ is the surface mass fraction of all metals, where the enrichment of CNO elements due to rotationally induced mixing is also taken into account.

Wolf-Rayet (WR) wind mass loss rates are calculated according to Hamann et al. (1995), but reduced by a factor of 10,

and including a metallicity dependence of $\dot{M} \propto Z_{\rm init}^{0.86}$ (Vink & de Koter 2005; see Fig. 1 of YL05):

$$\log\left(\frac{\dot{M}_{\rm WR}}{\rm M_\odot\,yr^{-1}}\right) = -12.95 + 1.5\log L/{\rm L_\odot} - 2.85X_{\rm s} \quad (1)$$
$$+0.86\log(Z_{\rm init}/Z_\odot)\ , \quad \text{for} \quad \log {\rm L/L_\odot} > 4.5,$$
$$= -36.8 + 6.8\log L/{\rm L_\odot} - 2.85X_{\rm s}$$
$$+0.86\log(Z_{\rm init}/Z_\odot)\ , \quad \text{for} \quad \log {\rm L/L_\odot} \leq 4.5\ .$$

As recent studies indicate that iron is likely the most important element for driving WR winds (Vink & de Koter 2005; Gräfener & Hamann 2005), YL05 and WH06 scaled the WR wind mass loss rates with the *initial* metallicity, as in the above equation. However, an enhancement of the CNO abundances at the surface to values higher than the initial metallicity (i.e. due to primary production) will also lead to enhanced WR winds, in particular at very low initial metallicity (e.g. Vink & de Koter 2005). As there exist no self-consistent WR wind models considering this effect, here we simply assume that the effective WR mass loss rate $\dot{M}^*_{\rm WR}$ increases linearly with the surface CNO abundance as follows:

$$\dot{M}^*_{\rm WR} = w_{\rm CNO} \cdot \dot{M}_{\rm WR} \quad (2)$$

where $\dot{M}_{\rm WR}$ is given by Eq. (2), and

$$w_{\rm CNO} = 1 + \max\left(19\frac{Z - Z_{\rm init}}{1 - Z_{\rm init}}, 0\right)\ . \quad (3)$$

Here $Z$ is the total mass fraction of all metals at the stellar surface. With this assumption, WC stars with $X_{\rm CNO} = 0.5$ and $X_{\rm He} = 0.5$ have an about 10 times higher mass loss rate than WN stars with $Z = Z_{\rm init}$. This choice is based on the results by Vink & de Koter (2005), which show that mass loss rates of WC stars with $X_{\rm C} = 0.5$ and $X_{\rm He} = 0.5$ are larger by an order of magnitude than those of WN stars when $X_{\rm Fe} \rightarrow 0$. However, this prescription cannot represent the non-linear behavior of WR mass loss rates as a function of surface abundance of heavy elements that is shown by Vink & de Koter, and must be regarded ad-hoc. Uncertainties due to this parameter are discussed in Sect. 3.2.

In the present study, we apply the above WR wind mass loss rates to stellar models with a surface helium mass fraction of $Y_{\rm s} \geq 0.7$, Kudritzki's mass loss rate for $Y_{\rm s} \leq 0.55$, and we interpolate between the two for $0.55 < Y_{\rm s} < 0.7$.

### 2.3. Core angular momentum threshold for GRBs

Within the collapsar scenario, the production of a GRB may be expected if those stars which undergo quasi-chemically homogeneous evolution retain enough angular momentum in the core. But it is currently uncertain exactly how much specific core angular momentum is required (MacFadyen & Woosley 1999; WH06; Lee & Ramirez-Ruiz 2006). Usually, the specific angular momentum for the last stable orbit around a black hole of a given mass (:= $j_{\rm LSO}$, Table 1; see Bardeen, Press & Teukolsky 1972) is adopted. The presence of magnetic fields may reduce the critical value by about 20% – 30% (Proga et al. 2003; Proga 2006, private communication).

**Table 1.** Specific angular momentum for the last stable orbit ($j_{\rm LSO}$) around a black hole with a given mass ($M_{\rm BH}$), calculated according to Bardeen, Press & Teukolsky (1972). Here, $j_{\rm SCH}$ and $j_{\rm Kerr,max}$ denote the cases for Schwarzschild black hole (non-rotating) and maximally rotating Kerr black hole, respectively.

| $M_{\rm BH}$ | $j_{\rm SCH}$ [cm$^2$ s$^{-1}$] | $j_{\rm Kerr,max}$ [cm$^2$ s$^{-1}$] |
|---|---|---|
| 2.0 $\rm M_\odot$ | $3.07 \times 10^{16}$ | $1.03 \times 10^{16}$ |
| 3.0 $\rm M_\odot$ | $4.60 \times 10^{16}$ | $1.55 \times 10^{16}$ |
| 4.0 $\rm M_\odot$ | $6.13 \times 10^{16}$ | $2.06 \times 10^{16}$ |
| 5.0 $\rm M_\odot$ | $7.66 \times 10^{16}$ | $2.57 \times 10^{16}$ |
| 10.0 $\rm M_\odot$ | $1.53 \times 10^{17}$ | $5.14 \times 10^{16}$ |

It is also uncertain whether GRB jets could be produced even if the innermost core ($\lesssim 2 - 3\ \rm M_\odot$) has a smaller angular momentum than $j_{\rm LSO}$, but when material further above has larger angular momentum (see WH06). Here, we assume that GRBs would be produced if any part of the CO core has a specific angular momentum larger than $j_{\rm LSO}$ (see, however, discussions in Sect. 4). Too large angular momenta ($j \gg 10^{17}$ cm$^2$s$^{-1}$ in the innermost region of $\sim 3\ \rm M_\odot$ may also prevent the formation of powerful jets (MacFadyen & Woosley 1999), but none of our models retains such large amounts of angular momentum in its core.

## 3. Uncertainties of the adopted physics

Note that the adopted physical assumptions in the present study differ from those in YL05 and WH06 in several respects. Firstly, we consider the effect of surface enrichment of CNO elements on WR winds as in Eqs. (2) and (3), while YL05 and WH06 do not. Secondly, we adopt faster semi-convection than in YL05 as explained above, but this new choice is comparable to the semi-convection efficiency employed by WH06. Thirdly, we consider the effect of the centrifugal force on the stellar structure as in YL05, but WH06 did not. However, the prescription of the angular momentum transport used in the present study is the same as that in YL05 and WH06.

As the pre-supernova stellar structure could be significantly affected by different physical assumptions, a good understanding of their influence on the stellar models is useful when applying our models to predict the cosmic GRB rate. In the following, we discuss the influence of the major assumptions on stellar evolution models.

### 3.1. Angular momentum transport

Our code employs the transport of angular momentum due to Eddington-Sweet circulations, shear instability, Goldreich-Schubert-Fricke instability, and magnetic torques according to the Spruit-Tayler dynamo, as explained in Heger et al. (2000) and Petrovic et al. (2005). The transport process is approximated as diffusion, as the full consideration of the interaction between the Eddington-Sweet circulation and the Spruit-Tayler dynamo (Maeder & Meynet 2005) requires solving a 4th-order differential equation, which is too expensive in computing time

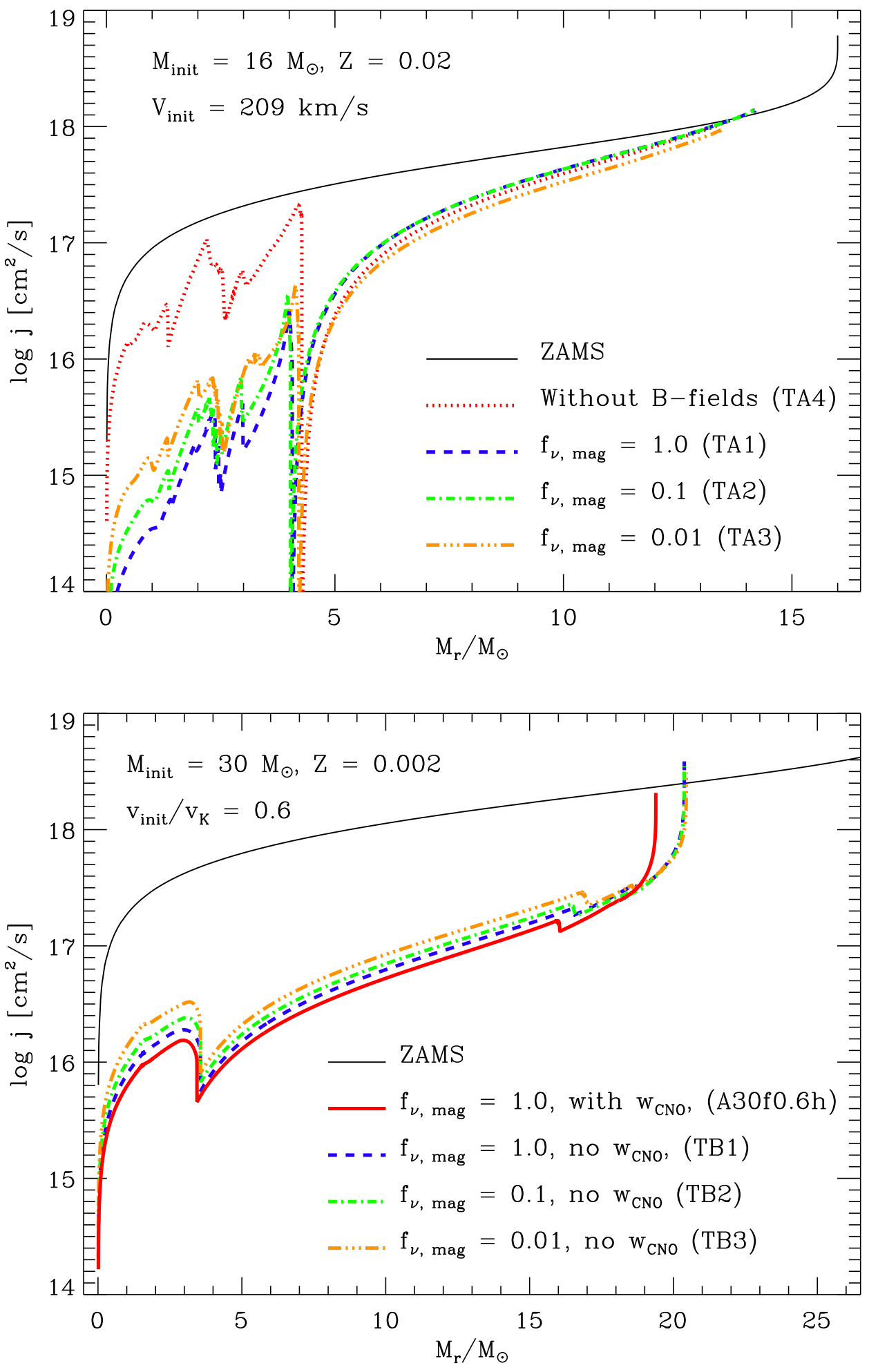

**Fig. 2.** *Upper panel*: Specific angular momentum as a function of the mass coordinate in stellar models of sequences TA1 (dashed), TA2 (dashed-dotted), TA3 (dashed-two-dotted), and TA4 (dotted), during core neon burning. The thin solid line corresponds to the specific angular momentum profile on the zero-age main sequence, which is the same for all cases. *Lower panel* Same as in the upper panel, but for sequences A30f0.3h (thick solid), TB1 (dashed), TB2 (dashed-dotted), and TB3 (three-dotted-dashed).

(G. Meynet, 2006, private communication). Under most circumstances, the magnetic torques dominate the internal transport of angular momentum. Although recent models including the Spruit-Tayler dynamo show that predicted spin rates of stellar remnants (neutron stars and white dwarfs) are consistent with observations (Heger, Woosley & Spruit 2005; Suijs et al. 2006), the order-of-magnitude estimate of the diffusive viscosity due to magnetic torques by Spruit (2002) might still be uncertain (Maeder & Meynet 2005; Spruit 2006).

To explore this uncertainty, we make simple experiments by introducing a free parameter $f_{\nu,\rm mag}$ such that

$$\nu_{\rm mag} = f_{\nu,\rm mag} \cdot \nu_{\rm ST}, \quad (4)$$

where $\nu_{\rm ST}$ is the magnetic viscosity according to Spruit (2002), and $\nu_{\rm mag}$ is the magnetic viscosity used in the code. Three different values of $f_{\nu,\rm mag}$ are used with two different initial models, as summarized in Table 2. Sequences TA1, TA2 and TA3 start with the same 16 $M_\odot$ ZAMS model with an initial equatorial rotation velocity of 30% of the Keplerian value ($v_{\rm init}/v_{\rm K}$ = 0.3) and $Z = 0.02$, with $f_{\nu,\rm mag}$ = 1.0, 0.1, and 0.01, respectively (see Fig. 2). Interestingly, a decrease of the magnetic viscosity by one order of magnitude leads to an increase of the specific angular momentum in the innermost 1.4$M_\odot$ by only a factor of two (compare TA1 and TA2; TA2 and TA3), and a decrease of $f_{\nu,\rm mag}$ by two orders of magnitude to an increase of $< j >_{1.4 M_\odot}$ by just a factor of four (compare TA1 and TA3), evaluated at core neon burning. This remarkable insensitivity of the core spin to $f_{\nu,\rm mag}$ is due to a self-regulation of the Spruit-Tayler dynamo: a smaller $f_{\nu,\rm mag}$ leads to a stronger degree of differential rotation, which in turn enhances the effective magnetic viscosity, and vice versa. Compared to non-magnetic models (Seq. TA4) — where the transport of angular momentum is dominated by Eddington-Sweet currents and shear instabilities (cf. Heger et al. 2000), magnetic models have less core angular momentum by more than one order of magnitude, for all considered values of $f_{\nu,\rm mag}$ (see also Fig. 2).

The influence of $f_{\nu,\rm mag}$ becomes even less important in that part of the parameter space where the CHES may produce GRBs, i.e. at low metallicity and rapid rotation. In sequences TB1, TB2 and TB3 ($M_{\rm init}$ = 30 $M_\odot$, $v_{\rm init}/v_{\rm K}$ = 0.6 and $Z = 0.002$), a decrease in $f_{\nu,\rm mag}$ by 100 results in a core angular momentum increase of only 80 % (compare TB1 and TB3 in Table 2; Fig. 2), evaluated during core oxygen burning. In these sequences, the stars undergo chemically homogeneous evolution, and the reduced sensitivity of the final core angular momentum to the parameter $f_{\nu,\rm mag}$ compared to sequences TA1–TA3 is due to the fact that any magnetic core-envelope coupling is rendered insignificant since the formation of a clear core-envelope structure is avoided altogether (see discussions in YL05).

### 3.2. Wolf-Rayet winds

We consider the effect of CNO surface enrichment on WR winds through the factor $w_{\rm CNO}$ in Eq. (2). A comparison of sequence TA1 (Table 2), where this effect is neglected, with sequence A30f0.6h (Table 5), shows that the inclusion of $w_{\rm CNO}$ leads to little changes in final mass and core angular momentum. However, its effect becomes larger for higher initial masses as revealed by comparing sequences TC1 and A40f0.6h: the CO core in sequence TC1 retains ten times more angular momentum in the CO core than in sequence A40f0.6h. Given that using the factor $w_{\rm CNO}$ according to Eq. (3) is rather ad-hoc, future systematic and self-consistent studies of the effect of surface abundance changes on the WR mass loss are highly desirable.

The adopted dependence of the WR wind mass loss rates on the WR star luminosity according to Hamann et al. (1995) may also need further investigation, as Hamann et al. did not consider the effect of clumping of WR winds. The effect of anisotropic mass loss from WR stars due to rotation (cf. Maeder & Meynet 2000) is another important factor to be carefully studied, as GRB progenitors must be rapidly rotating. Currently it is difficult to quantify the uncertainty due to these effects, and the WR winds remain as one of the most uncertain physics ingredients in our models.

### 3.3. The effect of the centrifugal force

Our models include the effect of the centrifugal force on the stellar structure following Endal & Sofia (1976; see also Meynet & Maeder 1997). Although this effect is not considered in WH06, it has non-negligible consequences in our GRB progenitor models. As indicated in Table 2, the 30 $M_\odot$ initial model at $v_{init}/v_K = 0.6$ has a higher equatorial rotational velocity and total angular momentum when the centrifugal force is neglected (Seq. TB4), compared to the case where it is considered (Seq. TB1), due to the change of the stellar structure and the corresponding adjustment of the moment of inertia.

Non-centrifugally-supported models are also more compact and hotter, leading to more efficient rotationally induced chemical mixing, and to less angular momentum loss for a given amount of mass loss. As a consequence, the core retains more angular momentum in sequences TB4 and TB6 than in sequence TB1.

More efficient mixing in sequences TB4 and TB6 also results in a smaller helium envelope during the WR phase than in sequence TB1. If the effect of surface enrichment of CNO elements on WR winds (i.e., $w_{CNO}$ in Eq. 2) is considered as in TB5 and TB7, the faster chemical mixing without the centrifugal effect leads to the loss of much more mass and angular momentum during WR phase than in the corresponding sequence A30f0.6h (see Table 5), where both, the centrifugal term and $w_{CNO}$, are included.

### 3.4. Semi-convection

The efficiency of semi-convective mixing in massive stars is currently not well constrained (e.g. Langer 1991). Model properties at $Z = 0.001$ with slow semi-convection ($\alpha_{SEM} = 0.04$) are presented in Table 3, and corresponding models with fast semi-convection ($\alpha_{SEM} = 1.0$) are shown in Table 6.

A comparison of the two cases reveals remarkable differences, in particular for the sequences which undergo chemically homogeneous evolution. As the use of slower semi-convection results in smaller CO cores and larger helium envelopes in the WR phase, the CO core is significantly more slowed down by the magnetic core-envelope coupling in this case, in particular for models with lower initial masses ($M_{init} \leq 20 M_\odot$). Therefore, the lower initial mass limit for GRB production shifts to higher initial masses when slower semi-convection is used.

At higher masses, on the other hand, less efficient mixing of CNO elements to the surface leads to the loss of less mass and angular momentum, and more angular momentum is retained in the core than in the corresponding cases with fast semi-convection (compare T30f0.4h, T30f0.5h and T30f0.6h with B30f0.4h and B30f0.5h), thus moving the upper limit for GRB production at $Z = 0.001$ to larger initial masses.

In conclusion, slower semi-convection shifts both the lower and upper initial mass limits for GRB production, to higher values: for $Z = 0.001$, the mass range of GRB production is $12 M_\odot < M_{init} \leq 30 M_\odot$ with $\alpha_{SEM} = 1.0$, and $20 M_\odot < M_{init} \leq 40 M_\odot$ with $\alpha_{SEM} = 0.04$. In addition, GRB progenitors have a more massive helium envelope for slower semi-convection, on average.

## 4. Model grids and the final fate of massive stars

With our fiducial assumptions described in Sect. 2, stellar model sequences are calculated, for various initial masses ($12 \lesssim M_{init}/M_\odot \lesssim 60$) and rotational velocities ($0.0 \lesssim v_{init}/v_K \lesssim 0.8$), and at 4 different metallicities ($Z = 0.004$, 0.002, 0.001 & 0.00001). Most sequences are followed until central carbon exhaustion or further. Model properties are presented in Tables 4 – 7. In those tables, the sequences which undergo the quasi-chemically homogeneous evolution are indicated with 'h' in the sequence number, while 'n' is the corresponding label for normal evolution. Here, a sequence is defined as evolving quasi-chemically homogeneously when the star becomes a WR star with $Y_s > 0.7$ during core hydrogen burning.

Based on the numerical results, we summarize the expected final fate of our models for each metallicity in the plane spanned by the initial mass and the initial fraction of the Keplerian value of the equatorial rotational velocity in Fig. 3.

As discussed by Maeder & Meynet (2000), the time scale of Eddington-Sweet circulations is proportional to Kelvin-Helmholtz time scale ($t_{KH}$) at a given ratio of the rotational velocity over the Keplerian velocity (i.e. $t_{ES} \propto t_{KH}(v_K/v_{rot})^2$). The Eddington-Sweet circulations are generally not spherically symmetric, and only their interaction with the baroclinic instability — which is essentially horizontal, and acts on the dynamical time scale — allows to pursue the computation of one-dimensonal rotating models (cf. Heger et al. 2000). I.e., the adoption of the isobaric surfaces as coordinate system implies an instant horizontal homogenisation of the chemical composition. A non-linear treatment of the interaction of the Eddington-Sweet circulations and the baroclinic instability with the magnetic fields is desirable, but not available yet (Maeder & Meynet 2005). This is to be kept in mind in the following discussion.

The Kelvin-Helmholtz time scale of main sequence stars decreases with increasing mass (Fig. 4). Although the hydrogen burning time ($t_{MS}$) also becomes smaller for higher initial masses, $t_{KH}$ usually decreases more rapidly than $t_{MS}$ in more massive stars, as shown in Fig. 4. This tendency is, in part, responsible for more efficient chemical mixing in higher mass stars (cf. Maeder & Meynet 2000). In addition, the entropy barrier becomes weakened in more massive stars due to the increased role of radiation pressure. These two effects result in more efficient mixing in higher mass stars, at a given $v_{init}/v_K$. This explains why the threshold value of $v_{init}/v_K$ for chemically homogeneous evolution [:= $(v_{init}/v_K)_{crit,CHEV}$] decreases with increasing initial mass, at a given metallicity.

Stars becomes significantly more compact as the metallicity becomes lower, while the change in luminosity is small.

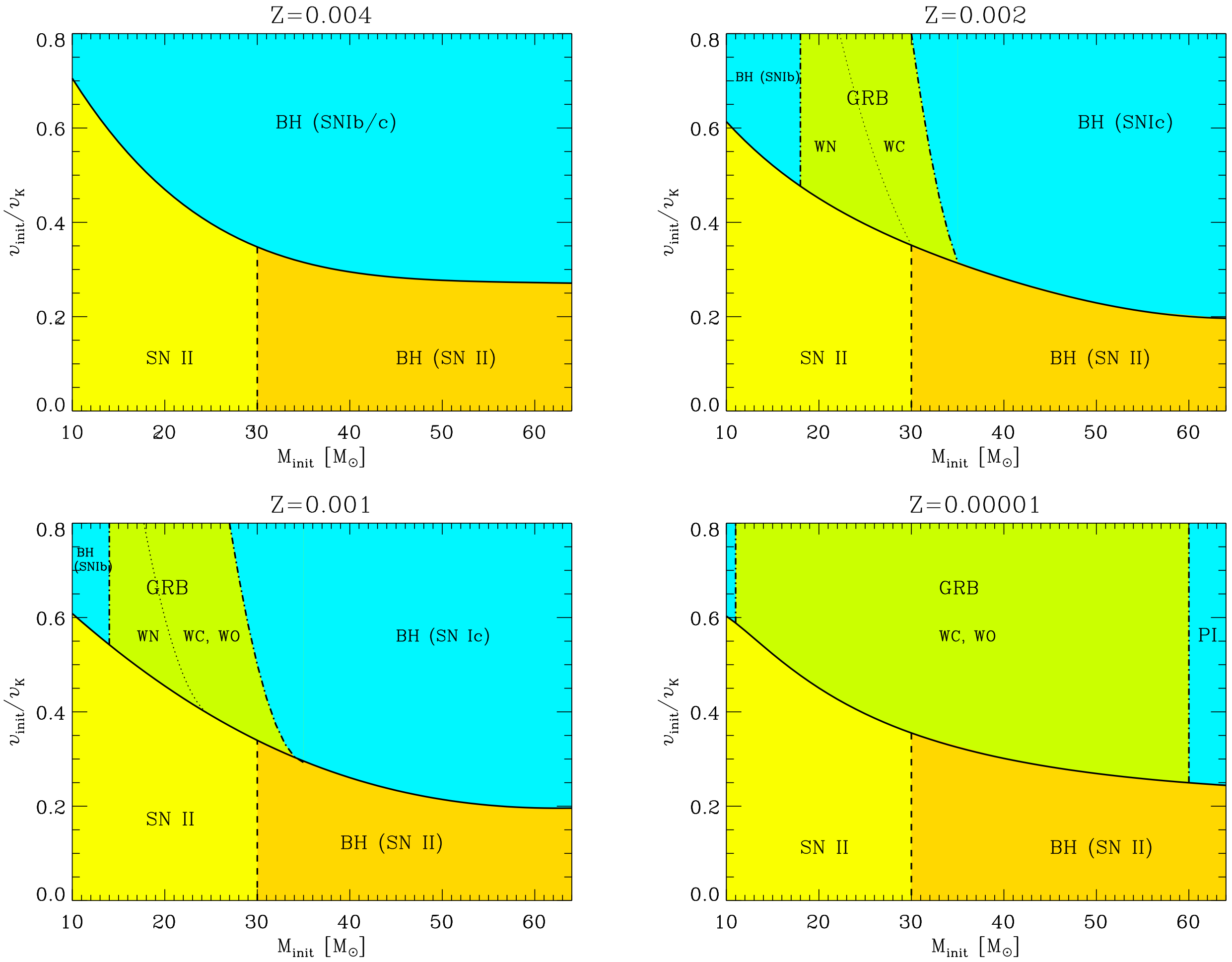

**Fig. 3.** Final fate of our rotating massive star models at four different metallicities ($Z$ = 0.004, 0.002, 0.001, & 0.00001), in the plane of initial mass and initial fraction of the Keplerian value of the equatorial rotational velocity. The solid line divides the plane into two parts, where stars evolve quasi-chemically homogeneous above the line, while they evolve into the classical core-envelope structure below the line. The dotted-dashed lines bracket the region of quasi-homogeneous evolution where the core mass, core spin and stellar radius are compatible with the collapsar model for GRB production (absent at Z=0.004). This GRB production region is divided into two parts, where GRB progenitors are WN or WC/WO types. To both sides of the GRB production region for $Z$ = 0.002 and 0.001, black holes are expected to form inside WR stars, but the core spin is insufficient to allow GRB production. For $Z$ = 0.00001, the pair-instability might occur to the right side of the GRB production region (see Heger et al 2003), although the rapid rotation may shift the pair instability region to larger masses. The dashed line in the region of non-homogeneous evolution separates Type II supernovae (SN II; left) and black hole (BH; right) formation, where the minimum mass for BH formation is simply assumed to be 30 $M_\odot$ (see, however, Heger et al. 2003 for a comprehensive discussion on the issue).

The thermal time thus increases with decreasing metallicity as shown in Fig. 4, and the diffusion coefficient for the chemical mixing by Eddington-Sweet circulations ($D_{\rm mix,ES}$) decreases accordingly. However, the value of $(v_{\rm init}/v_{\rm K})_{\rm crit,CHEV}$ remains to be nearly the same for all considered metallicities (Fig. 3), instead of increasing with decreasing metallicity. This can be ascribed to the following two factors. Firstly, the chemical mixing time itself ($\simeq R^2/D_{\rm mix}$) does not significantly change with decreasing metallicity, due to the reduced stellar size. Secondly, the spin-down effect due to stellar wind mass loss becomes more important in stars at higher metallicity, which tends to slow down the chemical mixing. The latter becomes particularly important at solar metallicity (Z=0.02), and $(v_{\rm init}/v_{\rm K})_{\rm crit,CHEV}$ largely increases compared to the subsolar metallicities considered in the present study (cf. YL05).

The regions of GRB production in Fig. 3 are determined according to the amount of angular momentum in the CO core of the corresponding models (cf. Sect. 2). In stars which undergo chemically homogeneous evolution, the CO core is spun down mainly by two factors: stellar wind mass loss and magnetic core-envelope coupling. At $Z$ = 0.004, angular momentum loss due to WR winds is so significant that the cores of the corresponding models retain only about 20% of the necessary angular momentum to produce a GRB. For lower metallicities ($Z$ = 0.002, 0.001, & 0.00001), the lower limit of the initial mass for GRB production is largely determined by the coupling between the helium envelope and the CO core by magnetic torques during the CO core contraction, as the ratio of the helium envelope mass to the CO core mass becomes larger for lower initial masses (see Tables 5 – 7). The upper limit of the initial mass for GRB production is mainly determined by the WR wind induced spin-down, for the cases of $Z = 0.001$ and 0.002. At $Z = 0.00001$, on the other hand, the rapidly rotating stars with $M \gtrsim 60\,{\rm M}_\odot$ form CO cores of $M \gtrsim 40\,{\rm M}_\odot$, which may be unstable to the pair instability (cf. Heger et al. 2003). The precise CO core mass limit for the pair instability must be a subject of future study, as it may increase with higher core angular momentum (Glatzel, Fricke & El Eid 1985).

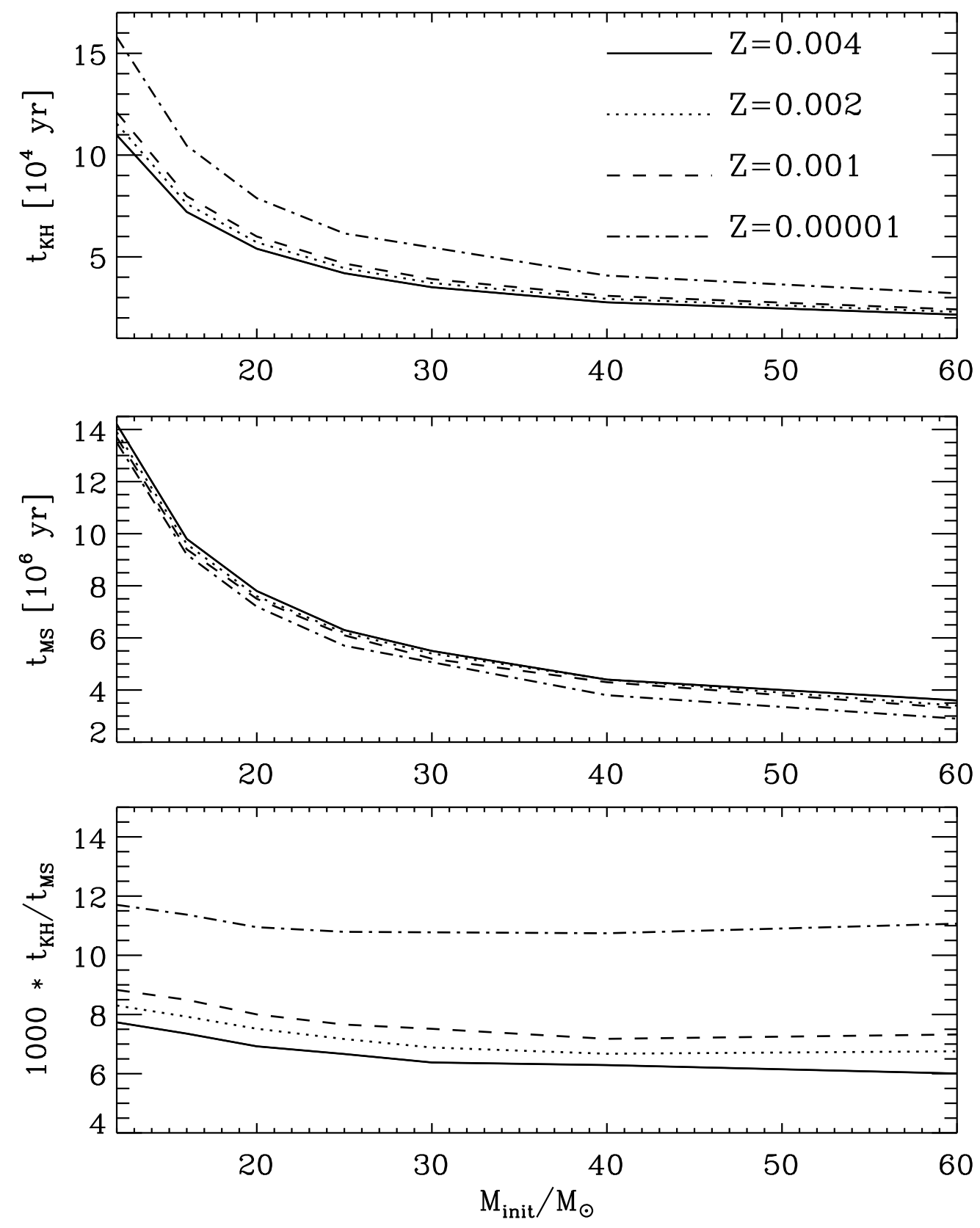

**Fig. 4.** *Top*: Kelvin-Helmholtz time ($t_{\rm KH}$) of non-rotating stars at zero-age main sequence as a function of initial mass, at different metallicities as indicated by the labels. *Middle*: Evolutionary time for core hydrogen burning ($t_{\rm MS}$) of non-rotating stars as a function of initial mass. *Bottom*: The ratio $t_{\rm KH}/t_{\rm MS}$ multiplied by 1000, as a function of initial mass

The prediction for GRB production in Fig. 3 is based on the assumption that GRBs are expected if any part of the CO core has a higher specific angular momentum than $j_{\rm LSO}$, as explained in Sect. 2. If we require instead that the innermost $2-3\,{\rm M}_\odot$ should have a specific angular momentum higher than $j_{\rm LSO}$, GRBs are expected only at $Z \lesssim 0.001$ according to our models. However, if the critical angular momentum for GRB production is reduced to about 80 % of $j_{\rm LSO}$, e.g., by the effect of magnetic fields, the expected GRB progenitor regions in Fig. 3 do not change significantly even if we only consider the innermost $2-3\,{\rm M}_\odot$ region.

The Wolf-Rayet types of GRB progenitors in Fig. 3 are determined according to the surface abundance of nitrogen, carbon and oxygen: WC (WN) stars are defined as WR stars with $X_{\rm N} < X_{\rm C}$ ($X_{\rm N} > X_{\rm C}$). We find that for our models, this criterion for WN and WC is comparable to that adopted by Eldridge & Vink (2006; WC if $X_{\rm C} + X_{\rm O} > 0.03$; see Tables 5 and 6). Some GRB progenitors at $Z = 0.001$ and 0.00001 are WO stars, which are defined as $Y_{\rm s} \leq X_{\rm C} + X_{\rm O}$ (see Tables 6 and 7; Eldridge & Vink 2006). Interestingly, some GRB progenitors are predicted to be WN stars with a rather thick helium envelope ($\Delta M_{\rm He} \approx 2.0\,{\rm M}_\odot$). Although a very high initial rotation velocity ($v_{\rm init}/v_{\rm K} \geq 0.4$) is required to produce such WN type GRB progenitors as they are mostly from relatively low mass stars ($M_{\rm init} < 25-30\,{\rm M}_\odot$; see Fig. 3), some supernovae accompanied by long GRBs are expected to be of Type Ib.

In Table 8, the evolution of core angular momentum and magnetic fields is illustrated for GRB progenitor models with $M_{\rm init} = 25\,{\rm M}_\odot$ at different metallicities. The numbers show a clear trend to stronger fields for lower metallicity, where the core rotation is faster. The obtained field strengths are up to two orders of magnitude larger than those obtained in the solar metallicity models of Heger et. al (2005). This might imply a stronger effect of magnetic fields in gamma-ray bursts at lower metallicity.

## 5. The GRB rate throughout the cosmic ages

Within the CHES, the fraction of massive stars which forms a long GRB depends on the distribution function of initial stellar rotation velocities, $D(v_{\rm init}/v_K)$. We derive $D(v_{\rm init}/v_K)$ from the stellar parameters of young O stars in the SMC cluster NGC 346 as measured by Mokiem et al. (2006; Fig. 5). NGC 346 is particularly suited due to its young age (2-4 Myr), and low (SMC) metallicity, which renders potential angular momentum loss due to O star winds unimportant. As protostellar winds could play an essential role in the formation of massive stars, $D(v_{\rm init}/v_{\rm K})$ might be a function of metallicity and stellar mass. However, in lack of better observational constraints, we assume it to be constant for all considered metallicities and masses.

Fig. 6 shows the predicted metallicity-dependent number ratio of GRBs versus core-collapse event ($:= f_{\rm GRB/SN}$), using different adopted distribution functions for $D(v_{\rm init}/v_{\rm K})$. Here we employ a Salpeter initial mass function (IMF). As also implied by Fig. 3, the GRB/SN ratio increases with decreasing metallicity. For the polynomial fits in the figure, we assumed $f_{\rm GRB/SN} \longrightarrow 0$ at $Z = 0.004$, as our models at $Z = 0.004$ have a too low spin to produce collapsars, while a significant number of GRBs is still expected at $Z = 0.002$. At very low metallicity ($Z \lesssim 10^{-5}$), the upper mass limit for GRB production is not determined by the spin-down due to stellar winds, but by the CO core mass beyond which the star is susceptible to the pair instability. Therefore, $f_{\rm GRB/SN}$ at $Z < 10^{-5}$ is not expected to significantly differ from $f_{\rm GRB/SN}$ at $Z = 10^{-5}$. We thus assumed $f_{\rm GRB/SN}$ to be constant at $Z \leq 10^{-5}$.

In Fig. 7, we show $f_{\rm GRB/SN}$ as a function of redshift, which is estimated using the gamma-fit for $D(v_{\rm init}/v_{\rm K})$ and the cosmic metallicity evolution model used by Langer & Norman (2006). Our model predicts $f_{\rm GRB/SN} \approx 8 \times 10^{-4}$ locally, and $f_{\rm GRB/SN} \approx 5 \times 10^{-3}$ globally, which are consistent with estimates based on observations (e.g. Podsiadlowski et al. 2004).

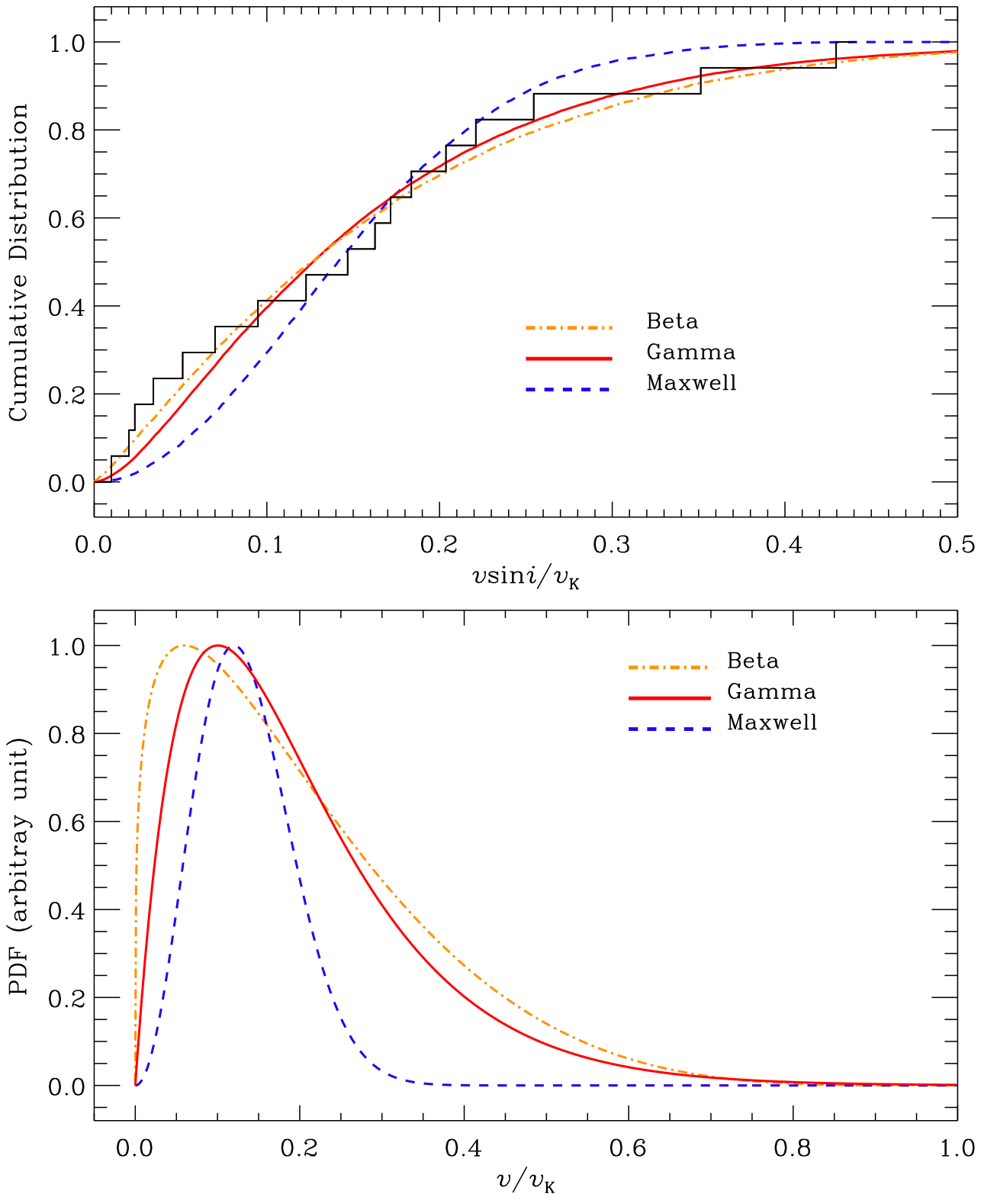

**Fig. 5.** *Upper panel*: Cumulative distribution of the fraction of the Keplerian value of the observed rotational velocity (i.e., $v \sin i$) of unevolved young stars in NGC 346 in small Magellanic clouds. The data are from Mokiem et al. (2006). The dotted-dashed, solid, and dashed lines are the best fits of synthesized distribution functions using three different distribution laws: beta, gamma and Maxwellian, respectively. Here we assume that stellar rotation axes are randomly oriented. *Lower panel*: The corresponding probability density function, as given by $P_{\rm Beta}(x; 0 \leq x \leq 1) = \frac{\Gamma(\alpha+\beta)}{\Gamma(\alpha)\Gamma(\beta)}(1-x)^{\beta-1}x^{\alpha-1}$ with $\alpha = 1.25$ and $\beta = 4.95$ (Beta distribution), $P_{\rm Gamma}(x; x \geq 0) = \frac{\lambda^{\nu}}{\Gamma(\nu)}x^{\nu-1}\exp(-\lambda x)$ with $\lambda = 9.95$ and $\nu = 2$ (Gamma distribution), and $P_{\rm Maxwell}(x; x \geq 0) = 4\pi x^2 (x_{\rm m})^{-3/2}\exp(-x^2/x_{\rm m}^2)$ with $v_{\rm m} = 0.1195$ (Maxwellian). Here $\Gamma(x)$ denotes the gamma function.

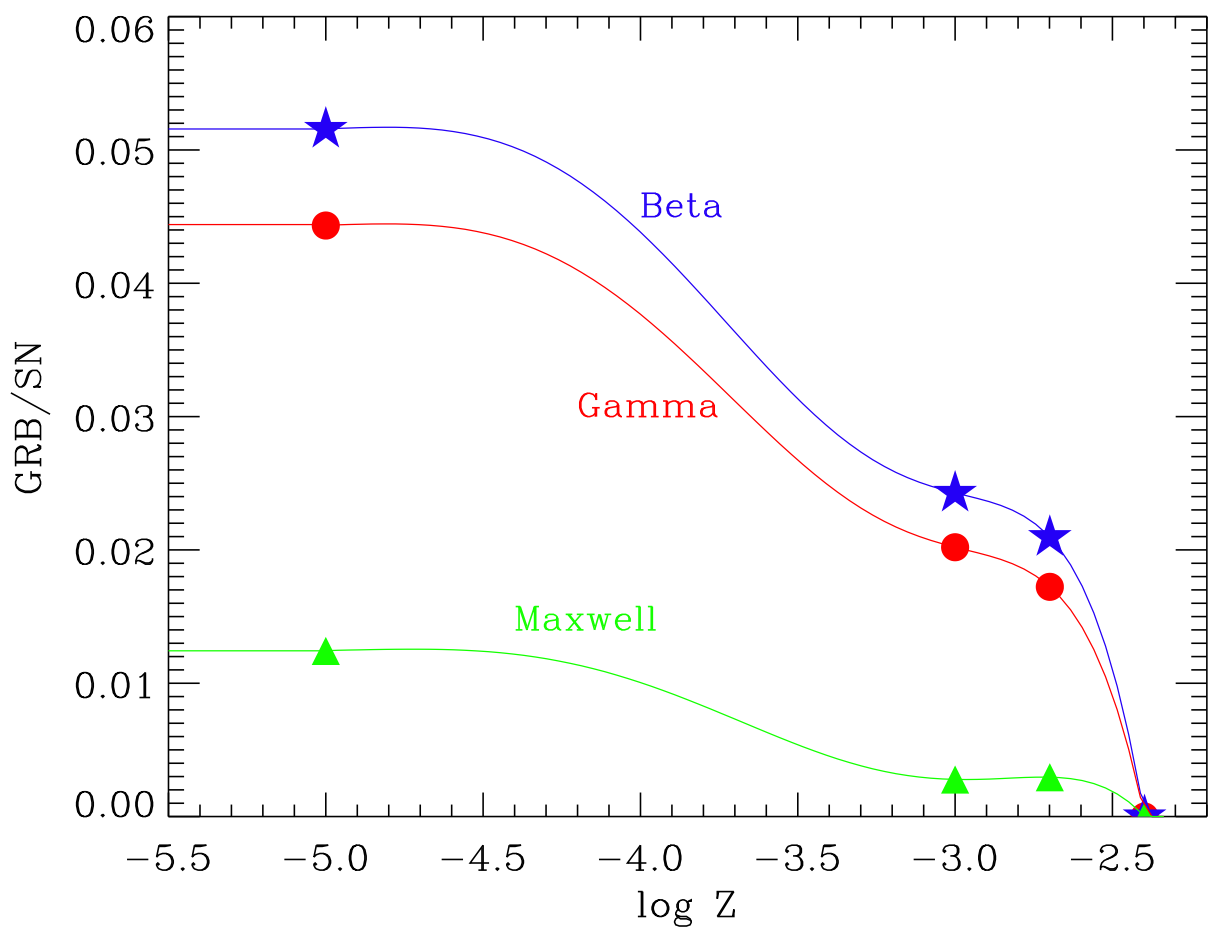

**Fig. 6.** The predicted number ratios of GRB progenitors over all massive stars ($8M_{\odot} < M < 100M_{\odot}$) as a function of metallicity, obtained by folding the three different adopted distributions of $v_{\rm init}/v_{\rm K}$ as given in Fig. 5 with the results of the stellar evolution grids as displayed in Fig. 3. The connecting lines are polynomial fits.

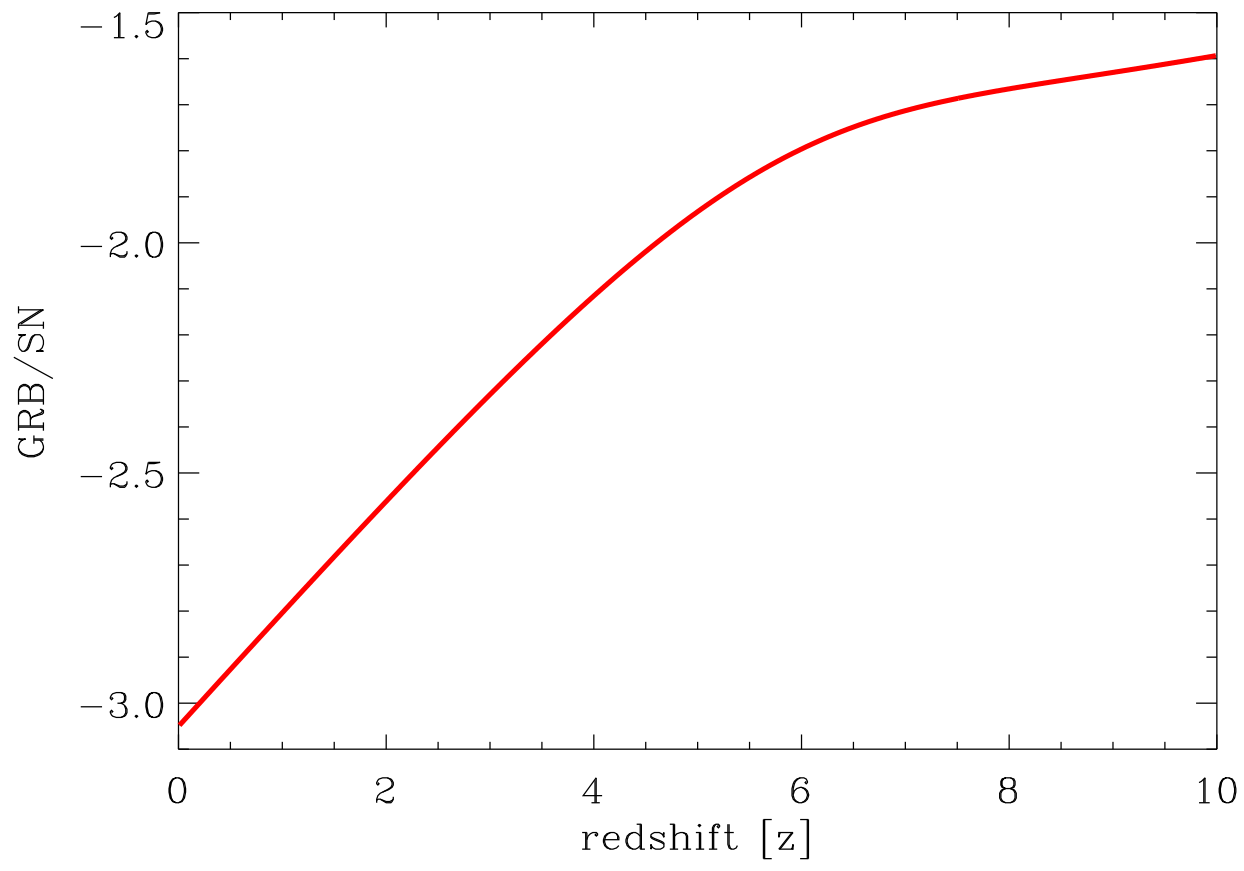

**Fig. 7.** Ratio of GRB versus core collapse supernova rate as a function of redshift, according to our GRB progenitor models. Note that the plotted ratio is independant of the adopted star formation history.

We also estimate the perceived GRB and core-collapse supernova rates in Fig. 8, following Langer & Norman (2006). Remarkably, comparison with the rate of core-collapse supernovae clearly indicates that the GRB rate according to the CHES does not follow the average cosmic star formation history (cf. Langer & Norman 2006). The observed SN and GRB rates are expected to peak at redshifts of $z \simeq 1.8$ and $z \simeq 2.8$, respectively. Our model also predicts a higher fraction of GRBs at high redshifts (i.e., more than 20% at $z > 6$; more than 50% at $z > 4$), than previous theoretical estimates (Natarajan et al. 2006; Bromm & Loeb 2006). This is mainly because GRB progenitors are limited to metallicities below $Z = 0.004$ according to our model.

In Fig. 9, we present the perceived GRB rate as a function of both metallicity and redshift. Assuming all GRBs are observable, the highest probability to find a GRB is located around $Z \approx 0.002$ and $z \approx 3$. A rather high probability to detect a GRB persists to the corner of $Z \simeq 10^{-3}$ and $z \gtrsim 6$, as our GRB progenitor models predict a higher GRB/SN-ratio for lower metallicity (Fig. 6). Future Swift observations will be a strong test to the predictions of our models (cf. Jakobsson et al. 2006).

## 6. Observational implications

The CHES predicts an evolution for metal-poor rapidly rotating massive stars which drastically differs from the commonly accepted evolutionary picture: Instead of forming an onion-skin structure and evolving to larger radii, stars avoid a chemical layering and only become more compact in the CHES evolution. Two things should be done before accepting the such extreme difference in evolution may indeed exist in nature: to elaborate all potential observable consequences, and then to rigorously test these predictions. In the preceding sections, we have worked out one of the most striking observational conse-quences of the CHES: the production of long GRBs, and their rate and metallicity at various redshifts. However, the CHES has more important implications.

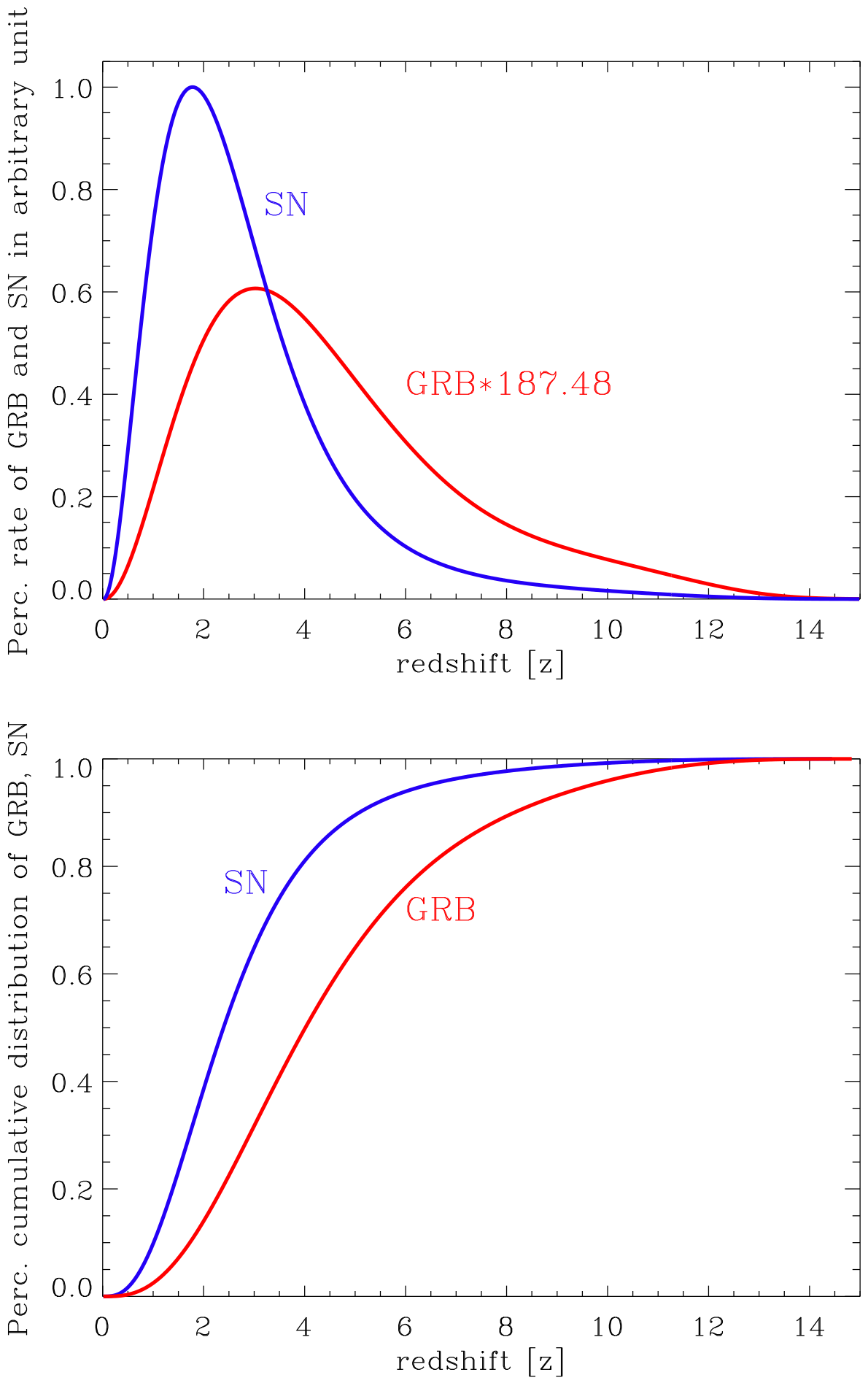

**Fig. 8.** *Upper panel*: Perceived supernova and GRB rate as function of redshift on an arbitrary scale, according to our GRB progenitor models, and for the specified cosmic metallicity evolution. The GRB rate is multiplied by a factor of 187.48, which is the perceived average ratio of SNe to GRBs in the universe, according to our models. *Lower panel*: Perceived cumulative number of SNe and GRBs as function of redshift. The GRB number has been multiplied by a factor of 187.48. The Y-scale is arbitrary but the same for both curves.

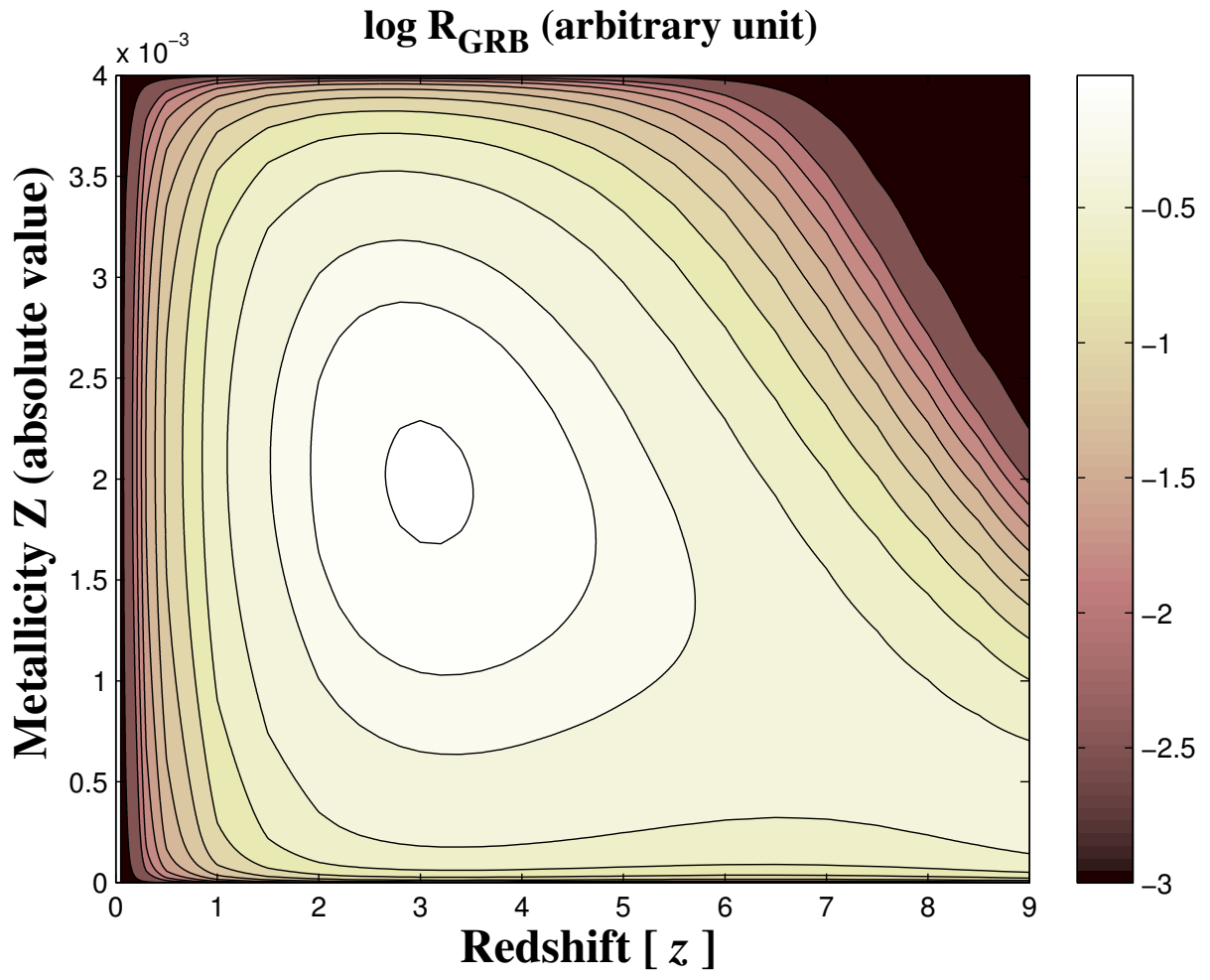

**Fig. 9.** Gray-shaded contour map (in color in the electronic version) of the perceived GRB rate with an arbitrary normalization, according to our GRB progenitor models and for the specified cosmic metallicity evolution, in the plane of redshift and metallicity.

## *6.1. GRB observations*

Concerning long GRBs, one of the most important prediction of the CHES is that of a strong bias of GRBs to low metallicities. While this is an inherent prediction of the collapsar scenario (see MacFadyen & Woosley 1999) due to the inescapable angular momentum loss by Wolf-Rayet winds which is stronger at higher metallicity (Vink & de Koter 2005), the CHES models allow to quantify this. Our present calculations imply that GRBs in the CHES frame should be restricted to metallicities below $Z \simeq 0.004$. It is important to point out this limit should apply to the abundance of iron, since this is the most important metal for wind driving in hot stars for not too small metallicities ($Z \gtrsim 0.0002$; Vink & de Koter 2005).

The estimated metallicity of some GRB host galaxies appears to be higher than this limit. For instance, the host galaxy of GRB980425 has a metallicity of $\sim$ 0.5 $Z_\odot$, which corresponds to $Z = 0.006 - 0.01$ depending on the value of $Z_\odot$ (although the region where GRB980425 actually occurred within this galaxy appears to have a lower metallicity; Hammer et al. 2006). However, as discussed in Sect. 3.2, the Wolf-Rayet mass loss rates constitute still a major uncertainty of the models. The discussion of the effect of wind clumping on these mass loss rates is ongoing, and a further reduction of the mass loss rates is far from excluded, which would result in an increase of the limiting metallicity for GRB production within the CHES. Future quantitative studies of stellar winds from rotating WR stars will be particularly important for better constraining the upper metallicity limit for single star GRB progenitors.

The CHES models from the grid provided in this paper also allow to predict the distribution of initial and final masses of GRB progenitors. With the distribution of rotational velocity given by Fig. 5, most GRB progenitors according to our models are predicted to have initial masses of about 25 $\sim$ 30 $M_\odot$ on average. This is significantly higher than average initial mass of core collapse supernovae, and may be relevant to the recent finding by Fruchter et al. (2006) that GRBs preferentially occur in brighter regions of their host galaxies than normal core collapse supernovae. The amount and distribution of angular momentum in the core of GRB progenitors, their final masses, their surface abundances, and their mass loss history — all things provided by the presented model grid (cf. Tables 5 ... 7) — have important implications for GRB observations (e.g., van Marle et al. 2005; 2006). For instance, GRB progenitors at lower metallicity should have, on average, higher angular momentum and stronger magnetic fields in the core, as discussed in Sect. 4, which might lead to more energetic GRBs at high redshifts. Our models also show that some GRB progenitors have a thick helium envelope, which may be associated with Type Ib supernovae, rather than Type Ic. We will discuss these important issues in a forthcoming paper in more detail (Cantiello, Yoon & Langer, in preparation).

### 6.2. Hypernovae

The models presented in Sect. 4 and in Tables 5 ... 7 predict a continuum of final core angular momenta, reaching from values of several times those required for collapsar formation down to values which are more than one order of magnitude below this (and are consistent with the spins of young neutron stars). Core angular momenta in the vicinity of, but below, the collapsar threshold may still cause considerable effects at the time of iron core collapse. It will be interesting to investigate which fraction of the CHES models are in that range, what their masses and metallicities are, which fractions form neutron stars and black holes, and to compare this with the observed properties of hypernovae (Nomoto et al. 2005).

### 6.3. Early universe

Our models with $Z = 0.00001$ imply that GRBs could also be produced abundantly from very metal poor populations, including the first stars in the universe, if some of them are born with large enough angular momentum. Although the probability to detect GRBs from Population III stars is limited by the limited number of Pop III stars (Wise & Abel 2005; Bromm & Loeb 2006), their progenitors may significantly affect the evolution of the early universe. The feature that the CHES models evolve to higher surface temperature already during core hydrogen burning, and have very high effective temperatures (up to 200 000 K; cf. YL05) later on. A detailed study of rapidly rotating Pop III stars in the context of the reionization of the universe may thus be a subject of interesting future work.

### 6.4. Chemical evolution

Quite obviously, the chemical yields of stars evolving within the CHES are quite different compared to the usual case. For instance, the metal cores of these stars are much larger compared to those of stars which evolve conventionally. However, additionally, the strong rotationally induced mixing triggers the formation of isotopes which are considered as secondary through primary nucleosynthesis. Most markedly, nitrogen can be enhanced by huge factors in this way, as demonstrated by the surface abundances of nitrogen displayed in Tables 5 ... 7. Recently, Chiappini et al. (2006) found strong primary nitrogen production in rotating very low metallicity massive star models. The models presented here may show an even stronger enhancement of the nitrogen yield. This may be related to observations of extremely metal-poor halo stars in our Galaxy, as well as nitrogen abundances in metal-poor galaxies.

### 6.5. Young star clusters

The obvious place for testing massive star evolution models is young star clusters. The problem with doing this for the CHES requires to find young clusters which have a low enough metallicity to allow for significant CHES effects. The only obvious cluster in this respect is NGC 346 in the SMC. Bouret et al. (2003) and Mokiem et al. (2006) find indications of the CHES to be realized for some stars in this cluster; however a more thorough investigation of this point appears worthwhile.

It should be noted that the CHES might not be irrelevant at larger metallicity. Although the increased main sequence winds disallow for chemically homogeneous evolution throughout, an incomplete CHES might apply: The most rapid rotators might undergo quasi-chemically homogeneous evolution for a fraction of their main sequence life, until the wind induced spin-down allows for the built-up of a chemical barrier inside the star. From that time on, the star would follow the standard evolutionary picture, i.e. form a core-envelope structure, but the envelope will the be already considerably enriched in CNO-burning products (cf. Maeder 1987; Langer 1992).

### 6.6. Metal-poor star forming galaxies

It is also interesting to note that our results presented in Fig. 3 suggest a different formation channel of WR stars at low metallicity. Traditionally, stellar wind mass loss has been regarded as the unique WR formation mechanism from single star progenitors. In particular, recent work by Meynet & Maeder (2005) indicates that in their rotating models without magnetic torques, the mass loss rate increases dramatically during the giant phase due to the surface enrichment of CNO materials induced by the shear instability, as a strong degree of differential rotation between the helium burning core and the hydrogen envelope persists. As a consequence, their models could predict a number ratio of WR to O stars consistent with observations, while non-rotating stellar models predicted too few WR stars, especially at low metallicity.

In our models with magnetic torques, strong shear mixing does not occur, as the degree of differential rotation is significantly weakened due to the magnetic core-envelope coupling. Therefore, if magnetic torques are important (cf. Sect. 1), CHES evolution may be the essential way to form WR stars at low metallicity. Our models at $Z = 0.004$ indicate that the WR/O ratio could reach a few percent at this metallicity for constant star formation, which is compatible with the observed WR/O ratio in SMC. Although the WR/O ratio may increase at higher metallicity due to the increased role of stellar winds, it is not expected to decrease for lower metallicity even down to $Z = 0.00001$, as most WR stars are produced through the CHES channel at $Z < 0.004$. This could explain observationally implied high WR/O ratios in metal poor WR galaxies (e.g. Fernandes et al. 2004; Crowther & Hadfield 2006) as well as in some metal-poor GRB host galaxies (Hammer et al. 2006).

## 7. Conclusions

Within the quasi-chemically-homogeneous evolution scenario (CHES) for GRB progenitors (YL05; WH06), we investigate the dependence of the GRB rate on metallicity and redshift, for the first time based on a grid of detailed massive star evo-lution models which include differential rotation and magnetic torques. We summarize our results as follows.

1. If the quasi-chemically homogeneous evolution scenario (CHES) provides the major channel for GRB production, most GRBs should occur at low metallicity. Our models predict a metallicity threshold of $Z \lesssim 0.004$, which is, however, subject to uncertain Wolf-Rayet mass loss rates. A reduction of the Wolf-Rayet mass loss rates, as currently discussed in the context of wind clumping, would lead to an increase of the metallicity threshold for GRB production through the CHES. Recent observations seem to imply that long GRBs may indeed prefer low metallicity environments, but quantitative comparisons are still difficult. A low-metallicity bias implies that GRBs are not an unbiased tracer of star formation (Fig. 8; Langer & Norman 2006).
2. The number ratio of GRBs to core-collapse supernovae as predicted by the CHES increases with decreasing metallicity (Figs. 3 and 6), as the Wolf-Rayet mass loss from quasi-chemically-homogeneously evolving stars becomes weaker. As a consequence, the CHES predicts a rather high GRB rate even at very low metallicity ($Z \lesssim 0.001$) and at high redshifts ($z > 6$; Figs. 8 and 9). For a standard cosmology ($\Omega_{\rm M} = 0.3, \Omega_\Lambda = 0.7$), GRBs at $Z \approx 0.002$ and $z \approx 3$ will be most commonly observed. These predictions need to be tested by future observations.
3. Our models predict that at least some supernovae associated with GRBs should be of Type Ib, which may be an interesting future test case for the CHES.
4. The CHES predicts a number ratio of GRBs versus core-collapse supernovae of about $8 \times 10^{-4}$ in the local universe, and about $5 \times 10^{-3}$ in an unbiased sample throughout the universe. These numbers may suffice to account for the observed number of GRBs, and may not require to invoke exotic binary evolutionary channels to produce long GRBs, as discussed by Langer & Norman (2006).
5. Before being accepted, the CHES needs to pass a number of observational tests (cf. Section 6), each of which deserves its own careful investigation. Those refer to properties of GRBs, their associated supernovae or circumstellar media, hypernova properties, chemical signatures in metal poor stars or galaxies, and the stellar content of young metal poor star clusters and metal poor star forming galaxies.

*Acknowledgements.* We wish to thank Rohied Mokiem and Alex de Koter for communicating their results prior to publication. We are grateful to Chris Fryer, George Meynet, Ken'ichi Nomoto, Daniel Proga, Philipp Podsiadlowski and Klaas Wiersema for helpful discussions. SCY is supported by the Netherlands Organization for Scientific Research (NWO) through the VENI grant (639.041.406).

## References

Bardeen, J.M., Press, W.H., & Teukolsky, S.A., 1972, ApJ, 178, 347
Bouret, J.-C., Lanz, T., & Hillier, D.J. et al., 2003, ApJ, 595, 1182
Bromm, V., & Loeb, A., 2006, ApJ, 642, 382
Chen, H.-W., Prochaska, J.X., Bloom, J.S., & Thomson, I.B., 2005, ApJ, 634, L25
Chiappini, C., Hirschi, R., Meynet, G., Ekstrm, S., Maeder, A., Matteucci, F., 2006, A&A, 449, L27
Conselice, C.J., Vreeswijk, P.M., & Fruchter, A.S., et al., 2005, ApJ, 633, 29
Crowther, P.A., & Hadfield, L.J., 2005, A&A, 449, 711
Gorosabel, J., Perez-Ramirez, D., & Sollerman, J., et al., 2005, A&A, 444, 711
Eggenberger, P., Maeder, A., & Meynet, G., 2005, A&A, 440, L9
Eldridge, J.J., & Vink, J.S., 2006, A&A, 452, 295
Endal, A.S., & Sofia, S., 1976, ApJ, 210, 184
Fernandes, I.F., de Carvalho, R., Contini, T., & Gal, R.R., 2004, MNRAS, 355, 728
Fruchter, A., Levan, A. J., Strolger, L., et al., 2006, Nature, 441, 463
Fryer, C.L., & Heger, A., 2005, ApJ, 623, 302
Fynbo, J.P.U., Jakobsson, P., & Möller, P., et al., 2003, A&A, 406, 63
Fynbo, J.P.U., Starling, R.L.C.., & Ledoux, C., et al., 2006, A&A, 451L, 47
Glatzel, W, Fricke, K.J., & El Eid, M.F., 1985, A&A, 149, 413
Gräfener, G., & Hamann, W.-R., 2005, A&A, 432, 633
Hamman, W.-R., Koesterke, L., & Wessolowski, U., 1995, A&A, 299, 151
Hammer, F., Floers, H., Schaerer, D. et al., 2006, A&A, submitted, [astro-ph/0604461]
Heger, A., Langer, N., & Woosley, S.E., 2000, A&A, 2000, ApJ, 528, 368
Heger, A., Fryer, C.L., Woosley, S.E., Langer, N., & Hartmann, D.H., 2003, ApJ, 591, 288
Heger, A., Woosley, S.E., & Spruit, H.C., 2005, ApJ, 626, 350
Hirschi, R., Meynet, G., & Maeder, A., 2005, A&A, 443, 581
Howarth, I.D., & Smith, K.C., 2001, MNRAS, 327, 353
Izzard, R.G., Ramirez-Ruiz, E., & C. A. Tout, 2004, MNRAS, 348, 1215
Jakobsson, P., Levan, A., Fynbo, J.P.U., et al, 2006, A&A, 447, 897
Kudritzki, R.P., Pauldrach, A., Puls, J., & Abbott, D.C., 1989, A&A, 219, 205
Langer, N., 1991, A&A, 252, 669
Langer, N., 1992, A&A, 265, L17
Langer, N., 1998, A&A, 329, 551
Langer, N., El Eid, M.F., & Fricke, K.J., 1985, A&A, 145, 179
Langer, N., & Norman, C., 2006, ApJL, 638, 63
MacFadyen, A.I., & Woosley, S.E., 1999, ApJ, 524, 262
Maeder, A., 1987, A&A, 178, 159
Maeder, A., & Meynet, G., 2000a, ARA&A, 38, 143
Maeder, A., & Meynet, G., 2000b, A&A, 361, 159
Maeder, A., & Meynet, G., 2004, A&A, 422, 225
Maeder, A., & Meynet, G., 2005, A&A, 440, 104
van Marle, A.-J., Langer, N., Garcia-Segura, G., 2005, A&A, 444, 837
van Marle, A.-J., Langer, N., Achterberg, A., Garcia-Segura, G., 2006, A&A, submitted, [astro-ph/0605698]
Meynet, G., & Maeder, A., 1997, A&A, 321, 465
Meynet, G., & Maeder, A., 2005, A&A, 429, 581
Mirabal, N., Halpern, J.P., An, D., Thorstensen, J.R., & Terndrup, D.M., 2006, ApJ, 643, L99
Mokiem, M.R., de Koter, A., C.J. Evans et al., A&A, 2006, in press
Natarajan, P., Albanna, B., Hjorth, J. et al., 2005, MNRAS
Nomoto, K., Maeda, K., Tominaga, N., Ohkubo, T., Deng, J., Mazzali, P.A.., 2005, Ap&SS, 298, 81
Ott, C., Burrows, A., Thompson, T.A., Livne, E., & Walder, R., 2006, ApJS, 164, 130
Petrovic, J., Langer, N., Yoon, S.-C., & Heger, A., 2005, A&A, 435, 247
Podsiadlowski, Ph., Mazzali, P.A., Nomoto, K., Lazzati, D., & Cappellaro, E., 2004, ApJ, 607, L17
Proga, D., MacFadyen, A.I., Armitage, Ph.J., & Beglman, M.C., ApJ, 599, 5

Lee, W.H., & Ramirez-Ruiz, E., 2006, ApJ, 641, 961
Spruit, H.C., 2002, A&A, 381, 923
Spruit, H.C., 2006, astro-ph/0607164
Stanek, K.Z., Gnedin, O.Y., & Beacom, J.F. et al., 2006, ApJ, submitted, [astro-ph/0604113]
Starling, R.L.C., Vreeswijk, P.M., & Ellison, S.L., et al., 2005, A&A, 442L, 21
Suijs, M., Langer, N., Yoon, S.-C., et al., 2006, A&A, in preparation
Vink, J.S., & de Koter, A., 2005, A&A, 442, 587
Vink, J.S., de Koter, A., & Lamers, H.J.G.L.M., 2001, A&A, 369, 574
Walborn, N., Morrell, N.I., Howarth, I.D., et. al, 2004, ApJ, 608, 1028
Wiersema, K., et al. 2006, A&A, to be submitted
Wise, J.H., & Abel, T., 2005, 629, 615
Woosley, S., 1993, ApJ, 405, 273
Woosley, S.E., & Heger, A., 2006, ApJ, 637, 914 (WH06)
Yoon, S.-C., & Langer, N., 2005, A&A, 443, 643 (YL05)
Yoon, S.-C., & Langer, N., 2006, to appear in Proc. "Stellar Evolution at Low Metallicity: Mass-Loss, Explosions, Cosmology" (eds: H. Lamers, N. Langer, & T. Nugis), ASP conf. Series, in press, [astro-ph/0511222]

**Table 2.** Model properties with test runs. Each column has the following meaning. $M_{\rm init}$: initial mass, $Z_{\rm init}$: initial absolute metallicity, $\rm v_{init}$: initial equatorial rotational velocity, $\rm v_{init}/v_K$: initial fraction of the Keplerian value of the equatorial rotational velocity. $J_{\rm init}$ : initial total angular momentum., $w_{\rm CNO}$ : consideration of the effect of the enrichment of CNO elements on the WR mass loss rate, $f_{\mu,\rm mag}$ : efficiency factor of the magnetic torque, $f_{\rm centri}$ : consideration of the effect of the centrifugal force on the stellar structure, End: the end point of the model sequence (YB: core helium burning, YE: central helium exhaustion, CB: core carbon burning, CE: central carbon exhaustion, NB: core neon burning, NE: central neon exhaustion, OB: core oxygen burning, OE: central oxygen exhaustion), $t_{\rm MS}$: evolutionary time from ZAMS to the end of main sequence, $t_{\rm f}$: evolutionary time from ZAMS to the end of the calculation, $M_{\rm f}$: final mass, $M_{\rm CO}$: CO core mass at the end of the calculation, $\Delta M_{\rm He}$: total helium mass in the envelope when the star ends as a WR star, $J_{\rm f}$: final angular momentum, $< j >_{\rm 3M_\odot}$: mean specific angular momentum of the innermost 1.4 $\rm M_\odot$, $< j >_{\rm 3M_\odot}$: mean specific angular momentum of the innermost 3 $\rm M_\odot$, $< j >_{\rm CO}$: mean specific angular momentum of the CO core.

| No. | $M_{\rm init}$ [$\rm M_\odot$] | $Z_{\rm init}$ | $\rm v_{init}$ [km/s] | $\rm v_{init}/v_K$ | $J_{\rm init}$ $10^{51}$ [erg/s] | $w_{\rm CNO}$ | $f_{\mu,\rm mag}$ | $f_{\rm centri}$ | End | $t_{\rm MS}$ $10^6$ [yr] | $t_{\rm f}$ $10^6$ [yr] | $M_{\rm f}$ [$\rm M_\odot$] | $M_{\rm CO}$ [$\rm M_\odot$] | $\Delta M_{\rm He}$ [$\rm M_\odot$] | $J_{\rm f}$ $10^{51}$ [erg/s] | $< j >_{\rm 1.4M_\odot}$ $10^{15}$ [$\rm cm^2/s$] | $< j >_{\rm 3M_\odot}$ $10^{15}$ [$\rm cm^2/s$] | $< j >_{\rm CO}$ $10^{15}$ [$\rm cm^2/s$] |
|---|---|---|---|---|---|---|---|---|---|---|---|---|---|---|---|---|---|---|
| TA1 | 16.00 | 0.020 | 209.53 | 0.30 | 20.70 | - | 1.00 | Yes | NB | 9.9 | 11.2 | 14.182 | 2.283 | - | 8.645 | 0.259 | 1.066 | 0.741 |
| TA2 | 16.00 | 0.020 | 209.53 | 0.30 | 20.70 | - | 0.10 | Yes | NB | 9.9 | 11.2 | 14.162 | 2.247 | - | 8.593 | 0.488 | 1.763 | 1.242 |
| TA3 | 16.00 | 0.020 | 209.53 | 0.30 | 20.70 | - | 0.01 | Yes | NB | 9.9 | 11.2 | 13.447 | 2.327 | - | 5.527 | 1.012 | 2.732 | 2.365 |
| TA4 | 16.00 | 0.020 | 209.53 | 0.30 | 20.70 | - | 0.00 | Yes | NB | 10.0 | 11.4 | 13.575 | 2.433 | - | 7.125 | 15.196 | 34.225 | 32.374 |
| TB1 | 30.00 | 0.002 | 522.80 | 0.60 | 127.94 | No | 1.00 | Yes | OB | 8.3 | 8.7 | 20.376 | 16.481 | 1.798 | 4.896 | 5.687 | 10.876 | 62.087 |
| TB2 | 30.00 | 0.002 | 522.80 | 0.60 | 127.94 | No | 0.10 | Yes | OB | 8.3 | 8.7 | 20.375 | 16.474 | 1.813 | 5.006 | 7.207 | 13.450 | 69.682 |
| TB3 | 30.00 | 0.002 | 522.80 | 0.60 | 127.94 | No | 0.01 | Yes | OB | 8.3 | 8.7 | 20.444 | 16.857 | 1.687 | 5.724 | 9.935 | 18.155 | 87.978 |
| TB4 | 30.00 | 0.002 | 549.56 | 0.60 | 147.51 | No | 1.00 | No | OB | 7.8 | 8.2 | 19.754 | 15.981 | 1.358 | 6.982 | 7.804 | 14.630 | 85.401 |
| TB5 | 30.00 | 0.002 | 549.56 | 0.60 | 147.51 | Yes | 1.00 | No | OB | 7.8 | 8.2 | 10.096 | 8.021 | 0.067 | 0.126 | 0.715 | 0.847 | 3.860 |
| TB6 | 30.00 | 0.002 | 457.97 | 0.50 | 122.93 | No | 1.00 | No | OE | 7.8 | 8.2 | 20.252 | 16.452 | 1.426 | 6.441 | 7.222 | 13.688 | 79.989 |
| TB7 | 30.00 | 0.002 | 457.97 | 0.50 | 122.93 | Yes | 1.00 | No | OB | 7.8 | 8.2 | 10.644 | 8.438 | 0.051 | 0.136 | 0.716 | 0.894 | 4.007 |
| TC1 | 40.00 | 0.002 | 555.94 | 0.60 | 216.20 | No | 1.00 | Yes | OE | 6.2 | 6.6 | 25.425 | 21.005 | 1.132 | 6.143 | 5.401 | 10.264 | 70.456 |

**Table 3.** Same as in Table 4, but with slow semi-convection ($\alpha_{\rm SEM}$ = 0.04) and with $Z$ = 0.001.

| No. | $M_{\rm init}$ [$M_\odot$] | $Z_{\rm init}$ | $v_{\rm init}$ [kms$^{-1}$] | $v_{\rm init}/v_{\rm K}$ | $J_{\rm init}$ $10^{51}$ [ergs$^{-1}$] | End | $t_{\rm MS}$ $10^6$ [yr] | $t_{\rm f}$ $10^6$ [yr] | $t_{\rm WR}$ $10^6$ [yr] | $M_{\rm f}$ [$M_\odot$] | $M_{\rm CO}$ [$M_\odot$] | $\Delta M_{\rm He}$ [$M_\odot$] | $Y_{\rm s}$ | $X_{\rm C}$ | $X_{\rm N}$ | $X_{\rm O}$ | $J_{\rm f}$ $10^{51}$ [erg/s] | $< j >_{3M_\odot}$ $10^{15}$ [cm$^2$/s] | $< j >_{\rm CO}$ $10^{15}$ [cm$^2$/s] |
|---|---|---|---|---|---|---|---|---|---|---|---|---|---|---|---|---|---|---|---|
| T12f0.5n | 12.0 | 0.001 | 377.59 | 0.50 | 19.73 | CB | 17.4 | 18.8 | 0.00 | 11.887 | 1.530 | - | 0.285 | -5.446 | -3.303 | -3.670 | 12.481 | 1.098 | 0.396 |
| T12f0.6h | 12.0 | 0.001 | 446.12 | 0.60 | 22.63 | OB | 27.2 | 27.6 | 3.03 | 10.670 | 3.201 | 5.920 | 0.911 | -5.087 | -3.173 | -4.930 | 1.714 | 0.987 | 1.135 |
| T12f0.7h | 12.0 | 0.001 | 513.14 | 0.70 | 25.35 | NB | 27.9 | 28.3 | 3.15 | 10.543 | 2.972 | 5.827 | 0.923 | -5.085 | -3.173 | -4.931 | 1.770 | 1.099 | 1.066 |
| T16f0.3n | 16.0 | 0.001 | 244.73 | 0.30 | 21.90 | CE | 9.8 | 10.7 | 0.00 | 15.912 | 1.870 | - | 0.250 | -4.924 | -3.498 | -3.390 | 19.202 | 0.776 | 0.415 |
| T16f0.4n | 16.0 | 0.001 | 323.45 | 0.40 | 28.46 | CE | 10.4 | 11.4 | 0.00 | 15.899 | 1.874 | - | 0.280 | -5.246 | -3.357 | -3.557 | 24.497 | 0.662 | 0.371 |
| T16f0.5h | 16.0 | 0.001 | 399.40 | 0.50 | 34.33 | NB | 17.0 | 17.3 | 2.24 | 13.934 | 4.838 | 7.129 | 0.910 | -5.070 | -3.173 | -4.954 | 2.310 | 0.958 | 2.322 |
| T16f0.6h | 16.0 | 0.001 | 471.84 | 0.60 | 39.37 | NE | 17.5 | 17.8 | 2.61 | 13.251 | 4.827 | 6.424 | 0.988 | -2.682 | -2.156 | -3.064 | 5.528 | 1.594 | 3.524 |
| T20f0.3n | 20.0 | 0.001 | 256.07 | 0.30 | 33.44 | CE | 7.9 | 8.5 | 0.00 | 19.850 | 2.238 | - | 0.281 | -4.842 | -3.417 | -3.483 | 28.202 | 0.654 | 0.546 |
| T20f0.4n | 20.0 | 0.001 | 338.40 | 0.40 | 43.45 | CE | 9.6 | 10.2 | 0.00 | 19.694 | 2.870 | - | 0.450 | -5.174 | -3.230 | -3.976 | 29.453 | 0.879 | 0.787 |
| T20f0.5h | 20.0 | 0.001 | 417.79 | 0.50 | 52.41 | CE | 12.6 | 12.9 | 2.05 | 16.511 | 7.131 | 7.321 | 0.961 | -5.021 | -3.174 | -4.952 | 4.331 | 1.698 | 5.096 |
| T20f0.8h | 20.0 | 0.001 | 640.86 | 0.80 | 74.48 | CE | 13.5 | 13.9 | 2.31 | 15.073 | 10.531 | 3.526 | 0.964 | -1.592 | -2.295 | -2.375 | 3.758 | 5.563 | 24.787 |
| T30f0.3n | 30.0 | 0.001 | 278.96 | 0.30 | 71.26 | NE | 5.6 | 6.1 | 0.00 | 29.437 | 5.213 | - | 0.367 | -5.040 | -3.270 | -3.788 | 39.577 | 0.779 | 1.804 |
| T30f0.4h | 30.0 | 0.001 | 368.54 | 0.40 | 92.56 | NE | 7.8 | 8.1 | 1.59 | 23.218 | 17.625 | 4.278 | 0.965 | -1.654 | -2.124 | -2.499 | 8.849 | 9.946 | 66.225 |
| T30f0.5h | 30.0 | 0.001 | 454.80 | 0.50 | 111.55 | NB | 7.9 | 8.3 | 1.68 | 21.184 | 17.443 | 1.291 | 0.894 | -1.075 | -2.171 | -1.888 | 5.795 | 12.340 | 80.872 |
| T40f0.4h | 40.0 | 0.001 | 392.21 | 0.40 | 156.73 | NB | 5.9 | 6.2 | 1.39 | 28.777 | 25.181 | 2.599 | 0.933 | -1.289 | -2.329 | -2.023 | 9.596 | 12.171 | 108.309 |
| T40f0.5h | 40.0 | 0.001 | 483.78 | 0.50 | 188.68 | NE | 6.0 | 6.4 | 1.47 | 17.060 | 14.151 | 0.070 | 0.062 | -0.484 | - | -0.216 | 0.476 | 2.028 | 9.414 |
| T60f0.3h | 60.0 | 0.001 | 324.55 | 0.30 | 250.30 | NE | 4.2 | 4.6 | 1.16 | 20.199 | 17.137 | 0.063 | 0.045 | -0.589 | - | -0.159 | 0.261 | 0.982 | 4.901 |

**Table 4.** Model properties with $Z$ = 0.004. Each column has the following meaning. $M_{\rm init}$: initial mass, $Z_{\rm init}$: initial absolute metallicity, $\rm v_{init}$: initial equatorial rotational velocity, $\rm v_{init}/v_K$: initial fraction of the Keplerian value of the equatorial rotational velocity. $J_{\rm init}$ : initial total angular momentum., End: the end point of the model sequence (YB: core helium burning, YE: central helium exhaustion, CB: core carbon burning, CE: central carbon exhaustion, NB: core neon burning, NE: central neon exhaustion, OB: core oxygen burning, OE: central oxygen exhaustion), $t_{\rm MS}$: evolutionary time from ZAMS to the end of main sequence, $t_{\rm f}$: evolutionary time from ZAMS to the end of the calculation, $t_{\rm WR}$: duration of WR stage, $M_{\rm f}$: final mass, $M_{\rm CO}$: CO core mass at the end of the calculation, $\Delta M_{\rm He}$: total helium mass in the envelope when the star ends as a WR star, $Y_{\rm s}$: surface helium mass fraction at the end of the calculation, $X_{\rm C}$, $X_{\rm N}$, $X_{\rm O}$: surface carbon, nitrogen and oxygen mass fraction in log scale at the end of the calculation, $J_{\rm f}$: final angular momentum, $< j >_{3\rm M_\odot}$: mean specific angular momentum of the innermost 3 $\rm M_\odot$, $< j >_{3\rm M_\odot}$: mean specific angular momentum of the CO core.

| No. | $M_{\rm init}$ [$\rm M_\odot$] | $Z_{\rm init}$ | $\rm v_{init}$ [$\rm kms^{-1}$] | $\rm v_{init}/v_K$ | $J_{\rm init}$ $10^{51}$ [$\rm ergs^{-1}$] | End | $t_{\rm MS}$ $10^6$ [yr] | $t_{\rm f}$ $10^6$ [yr] | $t_{\rm WR}$ $10^6$ [yr] | $M_{\rm f}$ [$\rm M_\odot$] | $M_{\rm CO}$ [$\rm M_\odot$] | $\Delta M_{\rm He}$ [$\rm M_\odot$] | $Y_{\rm s}$ | $X_{\rm C}$ | $X_{\rm N}$ | $X_{\rm O}$ | $J_{\rm f}$ $10^{51}$ [erg/s] | $< j >_{3\rm M_\odot}$ $10^{15}$ [$\rm cm^2/s$] | $< j >_{\rm CO}$ $10^{15}$ [$\rm cm^2/s$] |
|---|---|---|---|---|---|---|---|---|---|---|---|---|---|---|---|---|---|---|---|
| S12f0.5n | 12.0 | 0.004 | 354.14 | 0.50 | 19.34 | NB | 17.4 | 19.3 | 0.00 | 11.546 | 1.892 | - | 0.312 | -4.359 | -2.757 | -2.974 | 13.795 | 1.052 | 0.554 |
| S12f0.6n | 12.0 | 0.004 | 418.22 | 0.60 | 22.16 | YB | 23.8 | 24.4 | 0.00 | 11.446 | - | - | 0.455 | -4.736 | -2.590 | -3.732 | 11.075 | - | - |
| S12f0.7h | 12.0 | 0.004 | 480.82 | 0.70 | 24.79 | NB | 30.1 | 30.7 | 3.77 | 8.879 | 6.509 | 2.019 | 0.996 | -4.349 | -2.573 | -4.388 | 0.455 | 1.620 | 6.138 |
| S16f0.0n | 16.0 | 0.004 | 0.00 | 0.00 | 0.00 | NB | 9.8 | 11.1 | 0.00 | 15.727 | 2.677 | - | 0.289 | -3.398 | -3.105 | -2.769 | 0.000 | 0.000 | 0.000 |
| S16f0.1n | 16.0 | 0.004 | 77.60 | 0.10 | 7.43 | CB | 9.8 | 11.2 | 0.00 | 15.656 | 2.746 | - | 0.287 | -3.524 | -3.051 | -2.768 | 3.860 | 1.249 | 1.037 |
| S16f0.4n | 16.0 | 0.004 | 304.60 | 0.40 | 28.14 | NB | 10.9 | 12.2 | 0.00 | 15.000 | 2.959 | - | 0.297 | -4.261 | -2.804 | -2.900 | 6.648 | 1.099 | 1.066 |
| S16f0.5n | 16.0 | 0.004 | 375.95 | 0.50 | 33.90 | CE | 14.0 | 15.1 | 0.00 | 14.964 | 3.839 | - | 0.423 | -4.679 | -2.619 | -3.436 | 1.449 | 0.946 | 1.524 |
| S16f0.6h | 16.0 | 0.004 | 443.94 | 0.60 | 38.83 | CE | 18.8 | 19.3 | 2.85 | 11.110 | 8.499 | 2.031 | 0.996 | -4.157 | -2.579 | -4.438 | 0.469 | 2.604 | 8.565 |
| S20f0.0n | 20.0 | 0.004 | 0.00 | 0.00 | 0.00 | CB | 7.8 | 8.7 | 0.00 | 19.446 | 3.964 | - | 0.310 | -3.400 | -3.068 | -2.788 | 0.000 | 0.000 | 0.000 |
| S20f0.1n | 20.0 | 0.004 | 81.37 | 0.10 | 11.40 | CB | 7.8 | 8.7 | 0.00 | 19.352 | 4.018 | - | 0.311 | -3.528 | -3.013 | -2.791 | 3.513 | 0.950 | 1.626 |
| S20f0.2n | 20.0 | 0.004 | 162.15 | 0.20 | 22.56 | CB | 7.9 | 8.8 | 0.00 | 19.257 | 4.080 | - | 0.313 | -3.710 | -2.940 | -2.812 | 5.167 | 0.950 | 1.653 |
| S20f0.3n | 20.0 | 0.004 | 241.72 | 0.30 | 33.23 | NB | 8.6 | 9.5 | 0.00 | 19.027 | 4.329 | - | 0.347 | -4.024 | -2.804 | -2.919 | 3.729 | 0.939 | 1.739 |
| S20f0.4n | 20.0 | 0.004 | 319.34 | 0.40 | 43.15 | NB | 9.7 | 10.5 | 0.00 | 18.622 | 4.940 | - | 0.396 | -4.342 | -2.656 | -3.258 | 1.686 | 0.965 | 2.047 |
| S20f0.5h | 20.0 | 0.004 | 394.09 | 0.50 | 51.98 | CE | 13.5 | 13.9 | 2.32 | 13.385 | 10.638 | 1.951 | 0.996 | -4.133 | -2.580 | -4.456 | 0.492 | 2.298 | 10.117 |
| S20f0.6h | 20.0 | 0.004 | 465.28 | 0.60 | 59.52 | NB | 13.8 | 14.3 | 2.44 | 13.025 | 10.226 | 1.921 | 0.996 | -4.117 | -2.580 | -4.457 | 0.592 | 2.137 | 11.422 |
| S25f0.0n | 25.0 | 0.004 | 0.00 | 0.00 | 0.00 | CB | 6.3 | 7.0 | 0.00 | 23.834 | 5.904 | - | 0.344 | -3.426 | -2.991 | -2.833 | 0.000 | 0.000 | 0.000 |
| S25f0.2n | 25.0 | 0.004 | 170.17 | 0.20 | 34.39 | CB | 6.5 | 7.2 | 0.00 | 23.411 | 6.002 | - | 0.354 | -3.695 | -2.885 | -2.870 | 2.171 | 1.006 | 2.538 |
| S25f0.3n | 25.0 | 0.004 | 253.66 | 0.30 | 50.67 | CE | 7.5 | 8.2 | 0.00 | 22.063 | 7.048 | - | 0.417 | -4.071 | -2.732 | -3.048 | 1.339 | 1.057 | 3.029 |
| S25f0.4h | 25.0 | 0.004 | 335.05 | 0.40 | 65.77 | OE | 10.0 | 10.4 | 1.75 | 16.109 | 12.602 | 1.603 | 0.996 | -4.125 | -2.580 | -4.463 | 0.388 | 1.826 | 7.660 |
| S25f0.5h | 25.0 | 0.004 | 413.40 | 0.50 | 79.20 | OB | 10.3 | 10.7 | 1.97 | 15.436 | 12.254 | 1.503 | 0.996 | -4.004 | -2.581 | -4.443 | 0.663 | 2.441 | 12.458 |
| S25f0.6h | 25.0 | 0.004 | 487.95 | 0.60 | 90.63 | OE | 10.4 | 10.9 | 2.06 | 15.238 | 11.379 | 1.446 | 0.996 | -3.889 | -2.581 | -4.424 | 0.767 | 2.948 | 12.953 |
| S30f0.0n | 30.0 | 0.004 | 0.00 | 0.00 | 0.00 | CE | 5.5 | 6.0 | 0.00 | 27.430 | 8.013 | - | 0.356 | -3.426 | -2.972 | -2.849 | 0.000 | 0.000 | 0.000 |
| S30f0.2n | 30.0 | 0.004 | 177.11 | 0.20 | 48.34 | CE | 5.8 | 6.4 | 0.00 | 26.154 | 8.363 | - | 0.391 | -3.700 | -2.833 | -2.934 | 1.866 | 1.108 | 3.593 |
| S30f0.3n | 30.0 | 0.004 | 263.97 | 0.30 | 71.20 | CE | 7.0 | 7.5 | 0.00 | 21.254 | 11.107 | - | 0.600 | -4.279 | -2.618 | -3.500 | 0.718 | 1.166 | 4.606 |
| S30f0.4h | 30.0 | 0.004 | 348.61 | 0.40 | 92.38 | NB | 8.2 | 8.6 | 1.70 | 18.055 | 14.561 | 1.233 | 0.996 | -3.676 | -2.581 | -4.269 | 0.582 | 2.209 | 10.441 |
| S30f0.6h | 30.0 | 0.004 | 507.43 | 0.60 | 127.14 | OE | 8.5 | 8.9 | 1.87 | 17.332 | 13.819 | 1.132 | 0.992 | -2.374 | -2.588 | -3.391 | 0.902 | 3.283 | 15.464 |
| S40f0.0n | 40.0 | 0.004 | 0.00 | 0.00 | 0.00 | CE | 4.4 | 4.9 | 0.00 | 31.146 | 12.157 | - | 0.409 | -3.509 | -2.864 | -2.937 | 0.000 | 0.000 | 0.000 |
| S40f0.1n | 40.0 | 0.004 | 94.72 | 0.10 | 41.46 | CE | 4.5 | 4.9 | 0.00 | 30.139 | 12.182 | - | 0.427 | -3.588 | -2.823 | -2.974 | 1.936 | 1.202 | 5.080 |
| S40f0.2n | 40.0 | 0.004 | 188.72 | 0.20 | 82.04 | YB | 5.1 | 5.3 | 0.00 | 33.054 | - | - | 0.406 | -3.692 | -2.777 | -3.029 | 3.163 | - | - |
| S40f0.3h | 40.0 | 0.004 | 281.20 | 0.30 | 120.76 | YB | 6.1 | 6.1 | - | 32.093 | - | - | 0.885 | -4.323 | -2.575 | -4.344 | 22.964 | - | - |
| S40f0.4h | 40.0 | 0.004 | 371.22 | 0.40 | 156.55 | YB | 6.2 | 6.2 | - | 29.240 | - | - | 0.928 | -4.285 | -2.575 | -4.364 | 9.311 | - | - |
| S60f0.1n | 60.0 | 0.004 | 103.53 | 0.10 | 85.91 | YB | 3.6 | 3.6 | - | 53.059 | - | - | 0.300 | -3.334 | -3.048 | -2.827 | 14.370 | - | - |
| S60f0.2n | 60.0 | 0.004 | 206.21 | 0.20 | 169.85 | YB | 4.4 | 4.4 | - | 53.245 | - | - | 0.694 | -4.198 | -2.592 | -3.849 | 21.949 | - | - |
| S60f0.3h | 60.0 | 0.004 | 307.10 | 0.30 | 249.69 | YB | 4.4 | 4.4 | - | 43.724 | - | - | 0.900 | -4.297 | -2.575 | -4.368 | 26.856 | - | - |

**Table 5.** Same as in Table 4, but for $Z = 0.002$.

| No. | $M_{\rm init}$ [M$_\odot$] | $Z_{\rm init}$ | $\rm v_{init}$ [kms$^{-1}$] | $\rm v_{init}/v_K$ | $J_{\rm init}$ $10^{51}$ [ergs$^{-1}$] | End | $t_{\rm MS}$ $10^6$ [yr] | $t_{\rm f}$ $10^6$ [yr] | $t_{\rm WR}$ $10^6$ [yr] | $M_{\rm f}$ [M$_\odot$] | $M_{\rm CO}$ [M$_\odot$] | $\Delta M_{\rm He}$ [M$_\odot$] | $Y_{\rm s}$ | $X_{\rm C}$ | $X_{\rm N}$ | $X_{\rm O}$ | $J_{\rm f}$ $10^{51}$ [erg/s] | $< j >_{3M_\odot}$ $10^{15}$ [cm$^2$/s] | $< j >_{\rm CO}$ $10^{15}$ [cm$^2$/s] |
|---|---|---|---|---|---|---|---|---|---|---|---|---|---|---|---|---|---|---|---|
| A12f0.5n | 12.0 | 0.002 | 366.41 | 0.50 | 19.60 | YB | 17.2 | 18.0 | 0.00 | 11.852 | - | - | 0.286 | -4.918 | -3.022 | -3.330 | 17.292 | 2.187 | - |
| A12f0.55n | 12.0 | 0.002 | 399.97 | 0.55 | 21.09 | YB | 19.6 | 20.2 | 0.00 | 11.795 | - | - | 0.331 | -5.101 | -2.942 | -3.592 | 16.379 | 2.137 | - |
| A12f0.6h | 12.0 | 0.002 | 432.82 | 0.60 | 22.47 | OB | 28.3 | 28.9 | 3.45 | 9.667 | 7.207 | 2.162 | 0.994 | -4.655 | -2.872 | -4.672 | 1.426 | 2.918 | 14.531 |
| A12f0.8h | 12.0 | 0.002 | 561.88 | 0.80 | 27.79 | NB | 29.7 | 30.3 | 3.81 | 9.855 | 7.391 | 2.157 | 0.996 | -4.607 | -2.858 | -4.675 | 1.581 | 5.006 | 16.786 |
| A16f0.0n | 16.0 | 0.002 | 0.00 | 0.00 | 0.00 | NB | 9.6 | 10.9 | 0.00 | 15.861 | 2.700 | - | 0.275 | -3.700 | -3.431 | -3.056 | 0.000 | 0.000 | 0.000 |
| A16f0.3n | 16.0 | 0.002 | 237.95 | 0.30 | 21.85 | NB | 9.9 | 11.2 | 0.00 | 15.812 | 2.822 | - | 0.295 | -4.508 | -3.150 | -3.143 | 15.847 | 1.162 | 1.025 |
| A16f0.4n | 16.0 | 0.002 | 314.45 | 0.40 | 28.39 | NB | 10.7 | 11.9 | 0.00 | 15.738 | 2.918 | - | 0.323 | -4.857 | -3.046 | -3.279 | 15.779 | 1.154 | 1.077 |
| A16f0.5h | 16.0 | 0.002 | 388.21 | 0.50 | 34.24 | NE | 17.7 | 18.2 | 2.66 | 12.472 | 9.753 | 2.258 | 0.996 | -4.637 | -2.855 | -4.686 | 1.942 | 3.705 | 25.996 |
| A16f0.6h | 16.0 | 0.002 | 458.53 | 0.60 | 39.24 | NE | 18.1 | 18.6 | 2.75 | 12.277 | 9.596 | 2.252 | 0.997 | -4.570 | -2.808 | -4.617 | 2.126 | 3.873 | 27.603 |
| A16f0.8h | 16.0 | 0.002 | 595.30 | 0.80 | 48.56 | NE | 19.0 | 19.5 | 3.03 | 11.995 | 9.328 | 2.250 | 0.997 | -4.445 | -2.710 | -4.500 | 2.318 | 3.836 | 25.781 |
| A20f0.0n | 20.0 | 0.002 | 0.00 | 0.00 | 0.00 | NB | 7.6 | 8.5 | 0.00 | 19.739 | 3.940 | - | 0.292 | -3.695 | -3.401 | -3.073 | 0.000 | 0.000 | 0.000 |
| A20f0.3n | 20.0 | 0.002 | 249.20 | 0.30 | 33.45 | NB | 8.5 | 9.4 | 0.00 | 19.516 | 4.438 | - | 0.342 | -4.494 | -3.080 | -3.240 | 11.066 | 0.967 | 1.836 |
| A20f0.4n | 20.0 | 0.002 | 329.28 | 0.40 | 43.45 | NB | 9.1 | 9.9 | 0.00 | 18.921 | 5.188 | - | 0.349 | -4.693 | -2.969 | -3.501 | 2.751 | 0.968 | 2.155 |
| A20f0.45h | 20.0 | 0.002 | 368.30 | 0.45 | 48.07 | NE | 12.9 | 13.4 | 2.20 | 15.123 | 12.281 | 2.274 | 0.997 | -4.460 | -2.746 | -4.455 | 2.617 | 6.380 | 40.232 |
| A20f0.5h | 20.0 | 0.002 | 406.46 | 0.50 | 52.38 | OB | 13.1 | 13.5 | 2.19 | 14.962 | 12.239 | 2.275 | 0.997 | -4.272 | -2.662 | -4.308 | 2.811 | 6.413 | 42.906 |
| A20f0.6h | 20.0 | 0.002 | 480.01 | 0.60 | 60.02 | OB | 13.4 | 13.8 | 2.28 | 14.709 | 11.970 | 2.250 | 0.995 | -3.391 | -2.434 | -3.981 | 3.033 | 6.560 | 45.736 |
| A20f0.8h | 20.0 | 0.002 | 623.12 | 0.80 | 74.29 | OB | 13.9 | 14.4 | 2.51 | 14.350 | 11.536 | 2.252 | 0.995 | -3.245 | -2.411 | -3.896 | 3.269 | 6.753 | 47.338 |
| A25f0.0n | 25.0 | 0.002 | 0.00 | 0.00 | 0.00 | NB | 6.2 | 6.9 | 0.00 | 24.512 | 5.860 | - | 0.292 | -3.647 | -3.441 | -3.068 | 0.000 | 0.000 | 0.000 |
| A25f0.15n | 25.0 | 0.002 | 131.72 | 0.15 | 26.06 | CB | 6.3 | 7.0 | 0.00 | 24.399 | 5.843 | - | 0.306 | -3.921 | -3.273 | -3.103 | 8.513 | 1.013 | 2.460 |
| A25f0.3n | 25.0 | 0.002 | 261.28 | 0.30 | 50.91 | NB | 7.3 | 7.9 | 0.00 | 23.738 | 7.073 | - | 0.378 | -4.503 | -3.037 | -3.320 | 3.091 | 0.945 | 3.038 |
| A25f0.35n | 25.0 | 0.002 | 303.55 | 0.35 | 58.69 | YB | 8.7 | 9.2 | 0.00 | 22.850 | - | - | 0.514 | -4.852 | -2.931 | -3.660 | 1.249 | 1.578 | - |
| A25f0.4h | 25.0 | 0.002 | 345.18 | 0.40 | 66.12 | NB | 9.8 | 10.2 | 1.82 | 18.294 | 14.983 | 2.129 | 0.996 | -3.104 | -2.583 | -3.864 | 3.281 | 8.645 | 49.801 |
| A25f0.5h | 25.0 | 0.002 | 426.01 | 0.50 | 79.68 | OE | 10.0 | 10.4 | 1.88 | 17.828 | 14.549 | 2.014 | 0.992 | -2.466 | -2.453 | -3.481 | 3.674 | 9.445 | 52.555 |
| A25f0.6h | 25.0 | 0.002 | 502.95 | 0.60 | 91.25 | NB | 10.2 | 10.6 | 1.96 | 17.505 | 14.140 | 2.023 | 0.989 | -2.260 | -2.411 | -3.329 | 3.814 | 10.155 | 57.961 |
| A25f0.8h | 25.0 | 0.002 | 652.76 | 0.80 | 112.95 | OB | 10.6 | 11.0 | 2.08 | 16.894 | 13.504 | 1.809 | 0.978 | -1.811 | -2.364 | -3.017 | 3.670 | 9.205 | 52.003 |
| A30f0.3n | 30.0 | 0.002 | 271.77 | 0.30 | 71.48 | YB | 7.4 | 7.8 | 0.00 | 25.728 | - | - | 0.586 | -4.662 | -2.925 | -3.728 | 1.174 | 1.692 | - |
| A30f0.4h | 30.0 | 0.002 | 358.99 | 0.40 | 92.80 | OB | 8.0 | 8.4 | 1.64 | 21.003 | 17.285 | 1.736 | 0.989 | -2.182 | -2.587 | -3.231 | 3.852 | 8.654 | 51.633 |
| A30f0.5h | 30.0 | 0.002 | 442.94 | 0.50 | 111.78 | OB | 8.2 | 8.5 | 1.69 | 20.383 | 16.574 | 1.582 | 0.982 | -1.889 | -2.485 | -2.930 | 4.065 | 9.489 | 54.373 |
| A30f0.6h | 30.0 | 0.002 | 522.80 | 0.60 | 127.94 | OB | 8.3 | 8.7 | 1.76 | 19.387 | 15.974 | 1.061 | 0.940 | -1.300 | -2.475 | -2.268 | 3.280 | 8.636 | 49.101 |
| A40f0.0n | 40.0 | 0.002 | 0.00 | 0.00 | 0.00 | CB | 4.4 | 4.8 | 0.00 | 36.743 | 12.318 | - | 0.318 | -3.659 | -3.356 | -3.112 | 0.000 | 0.000 | 0.000 |
| A40f0.2n | 40.0 | 0.002 | 194.20 | 0.20 | 82.30 | CB | 4.8 | 5.3 | 0.00 | 35.061 | 13.274 | - | 0.379 | -4.026 | -3.086 | -3.305 | 3.053 | 1.222 | 5.468 |
| A40f0.25h | 40.0 | 0.002 | 242.05 | 0.25 | 102.05 | OB | 6.0 | 6.4 | 0.90 | 29.346 | 25.320 | 2.159 | 0.998 | -4.434 | -2.880 | -4.788 | 1.240 | 2.098 | 15.273 |
| A40f0.3h | 40.0 | 0.002 | 289.42 | 0.30 | 121.22 | NB | 6.1 | 6.5 | 1.38 | 27.275 | 23.427 | 1.213 | 0.994 | -2.433 | -2.865 | -3.351 | 3.957 | 6.326 | 48.502 |
| A40f0.4h | 40.0 | 0.002 | 382.17 | 0.40 | 157.26 | OB | 6.2 | 6.6 | 1.49 | 26.322 | 22.156 | 1.032 | 0.977 | -1.753 | -2.749 | -2.584 | 4.746 | 7.852 | 56.497 |
| A40f0.6h | 40.0 | 0.002 | 555.94 | 0.60 | 216.20 | OB | 6.4 | 6.8 | 1.63 | 15.371 | 14.667 | 0.028 | 0.033 | -0.569 | - | -0.159 | 0.241 | 1.340 | 6.908 |
| A60f0.1n | 60.0 | 0.002 | 106.57 | 0.10 | 86.31 | YB | 3.5 | 3.6 | 0.00 | 54.809 | - | - | 0.294 | -3.649 | -3.332 | -3.136 | 12.270 | 2.250 | - |
| A60f0.2h | 60.0 | 0.002 | 212.30 | 0.20 | 170.71 | YB | 4.3 | 4.4 | 0.33 | 49.932 | - | - | 0.855 | -4.566 | -2.879 | -4.574 | 4.136 | 2.236 | - |
| A60f0.3h | 60.0 | 0.002 | 316.24 | 0.30 | 251.16 | NB | 4.3 | 4.7 | 1.19 | 13.652 | 13.247 | 0.032 | 0.043 | -0.564 | - | -0.167 | 0.097 | 0.749 | 3.354 |

**Table 6.** Same as in Table 4, but for $Z = 0.001$.

| No. | $M_{\rm init}$ | $Z_{\rm init}$ | $v_{\rm init}$ | $v_{\rm init}/v_{\rm K}$ | $J_{\rm init}$ | End | $t_{\rm MS}$ | $t_{\rm f}$ | $t_{\rm WR}$ | $M_{\rm f}$ | $M_{\rm CO}$ | $\Delta M_{\rm He}$ | $Y_{\rm s}$ | $X_{\rm C}$ | $X_{\rm N}$ | $X_{\rm O}$ | $J_{\rm f}$ | $< j >_{3{\rm M}_\odot}$ | $< j >_{\rm CO}$ |
|---|---|---|---|---|---|---|---|---|---|---|---|---|---|---|---|---|---|---|---|
| | | | | | $10^{51}$ | | $10^6$ | $10^6$ | $10^6$ | | | | | | | | $10^{51}$ | $10^{15}$ | $10^{15}$ |
| | $[{\rm M}_\odot]$ | | $[{\rm kms}^{-1}]$ | | $[{\rm ergs}^{-1}]$ | | [yr] | [yr] | [yr] | $[{\rm M}_\odot]$ | $[{\rm M}_\odot]$ | $[{\rm M}_\odot]$ | | | | | [erg/s] | $[{\rm cm}^2/{\rm s}]$ | $[{\rm cm}^2/{\rm s}]$ |
| B12f0.0n | 12.0 | 0.001 | 0.00 | 0.00 | 0.00 | CB | 13.7 | 15.8 | 0.00 | 11.964 | 1.748 | - | 0.257 | -4.039 | -3.734 | -3.346 | 0.000 | 0.000 | 0.000 |
| B12f0.3n | 12.0 | 0.001 | 231.33 | 0.30 | 12.59 | NB | 14.2 | 16.2 | 0.00 | 11.953 | 1.823 | - | 0.270 | -4.962 | -3.468 | -3.419 | 11.619 | 1.188 | 0.522 |
| B12f0.5n | 12.0 | 0.001 | 377.59 | 0.50 | 19.73 | NB | 17.1 | 18.8 | 0.00 | 11.887 | 2.156 | - | 0.303 | -5.392 | -3.282 | -3.733 | 13.785 | 1.047 | 0.683 |
| B12f0.6h | 12.0 | 0.001 | 446.12 | 0.60 | 22.63 | NE | 27.5 | 28.0 | 3.52 | 9.989 | 7.494 | 2.136 | 0.997 | -4.690 | -3.031 | -4.780 | 1.731 | 5.520 | 18.697 |
| B12f0.8h | 12.0 | 0.001 | 579.43 | 0.80 | 28.02 | CB | 28.8 | 29.4 | 3.80 | 10.121 | 7.642 | 2.084 | 0.997 | -4.201 | -2.649 | -4.220 | 1.850 | 6.455 | 21.506 |
| B16f0.0n | 16.0 | 0.001 | 0.00 | 0.00 | 0.00 | CB | 9.4 | 10.7 | 0.00 | 15.923 | 2.733 | - | 0.256 | -3.963 | -3.828 | -3.330 | 0.000 | 0.000 | 0.000 |
| B16f0.4n | 16.0 | 0.001 | 323.45 | 0.40 | 28.46 | NB | 10.5 | 11.7 | 0.00 | 15.845 | 2.940 | - | 0.320 | -5.287 | -3.331 | -3.605 | 20.892 | 1.137 | 1.087 |
| B16f0.45n | 16.0 | 0.001 | 361.84 | 0.45 | 31.49 | CE | 13.3 | 14.3 | 0.00 | 15.622 | 4.071 | - | 0.430 | -5.308 | -3.234 | -3.938 | 10.838 | 0.967 | 1.662 |
| B16f0.5h | 16.0 | 0.001 | 399.40 | 0.50 | 34.33 | OB | 17.1 | 17.6 | 2.56 | 13.074 | 10.381 | 2.299 | 0.996 | -4.110 | -2.467 | -3.960 | 3.564 | 5.734 | 44.971 |
| B16f0.6h | 16.0 | 0.001 | 471.84 | 0.60 | 39.37 | OB | 17.5 | 18.0 | 2.63 | 12.795 | 10.156 | 2.243 | 0.994 | -3.745 | -2.260 | -3.722 | 3.480 | 5.360 | 44.027 |
| B16f0.8h | 16.0 | 0.001 | 612.84 | 0.80 | 48.79 | NB | 18.3 | 18.8 | 2.78 | 12.443 | 9.851 | 2.170 | 0.986 | -2.299 | -2.160 | -3.344 | 3.328 | 6.568 | 43.474 |
| B20f0.0n | 20.0 | 0.001 | 0.00 | 0.00 | 0.00 | CB | 7.5 | 8.4 | 0.00 | 19.858 | 4.072 | - | 0.259 | -3.914 | -3.873 | -3.330 | 0.000 | 0.000 | 0.000 |
| B20f0.3n | 20.0 | 0.001 | 256.07 | 0.30 | 33.44 | NB | 8.6 | 9.5 | 0.00 | 19.707 | 4.639 | - | 0.359 | -4.999 | -3.338 | -3.604 | 17.837 | 0.971 | 1.921 |
| B20f0.45h | 20.0 | 0.001 | 378.53 | 0.45 | 48.09 | OB | 12.5 | 12.9 | 2.06 | 16.068 | 13.301 | 2.380 | 0.990 | -3.092 | -2.109 | -3.425 | 5.189 | 10.110 | 72.383 |
| B20f0.5h | 20.0 | 0.001 | 417.79 | 0.50 | 52.41 | NE | 12.6 | 13.1 | 2.20 | 15.805 | 12.991 | 2.311 | 0.981 | -2.079 | -2.070 | -3.061 | 5.106 | 10.230 | 73.584 |
| B20f0.7h | 20.0 | 0.001 | 567.52 | 0.70 | 67.32 | NB | 13.2 | 13.6 | 2.28 | 15.028 | 12.169 | 2.087 | 0.943 | -1.352 | -2.180 | -2.458 | 4.625 | 11.522 | 69.314 |
| B20f0.8h | 20.0 | 0.001 | 640.86 | 0.80 | 74.48 | NB | 13.5 | 13.9 | 2.38 | 14.790 | 12.123 | 2.012 | 0.933 | -1.267 | -2.194 | -2.356 | 4.492 | 10.172 | 69.383 |
| B25f0.0n | 25.0 | 0.001 | 0.00 | 0.00 | 0.00 | CB | 6.1 | 6.8 | 0.00 | 24.742 | 5.932 | - | 0.249 | -3.810 | -4.087 | -3.313 | 0.000 | 0.000 | 0.000 |
| B25f0.3n | 25.0 | 0.001 | 268.30 | 0.30 | 50.81 | CE | 7.3 | 7.9 | 0.00 | 24.312 | 7.216 | - | 0.378 | -5.002 | -3.291 | -3.721 | 9.541 | 1.059 | 3.094 |
| B25f0.4h | 25.0 | 0.001 | 354.51 | 0.40 | 66.02 | NB | 9.4 | 9.8 | 1.84 | 19.325 | 15.684 | 2.048 | 0.947 | -1.394 | -2.150 | -2.418 | 6.110 | 13.806 | 85.338 |
| B25f0.5h | 25.0 | 0.001 | 437.58 | 0.50 | 79.59 | OE | 9.6 | 10.0 | 1.86 | 18.302 | 13.229 | 1.317 | 0.905 | -1.111 | -2.190 | -2.060 | 5.142 | 12.113 | 48.696 |
| B25f0.6h | 25.0 | 0.001 | 516.71 | 0.60 | 91.20 | NB | 9.8 | 10.2 | 1.99 | 16.569 | 13.471 | 0.554 | 0.688 | -0.621 | -2.442 | -1.172 | 2.904 | 9.669 | 51.205 |
| B25f0.7h | 25.0 | 0.001 | 594.13 | 0.70 | 102.16 | NB | 10.0 | 10.4 | 2.08 | 15.962 | 12.933 | 0.413 | 0.516 | -0.474 | -2.647 | -0.842 | 2.520 | 8.666 | 46.156 |
| B25f0.8h | 25.0 | 0.001 | 670.86 | 0.80 | 113.04 | NE | 10.2 | 10.7 | 2.02 | 15.653 | 12.693 | 0.391 | 0.484 | -0.452 | -2.752 | -0.799 | 2.399 | 8.018 | 44.721 |
| B25f0.9h | 25.0 | 0.001 | 747.30 | 0.90 | 124.06 | NB | 10.5 | 10.9 | 2.03 | 15.445 | 12.477 | 0.396 | 0.505 | -0.462 | -2.725 | -0.834 | 2.361 | 8.287 | 44.185 |
| B30f0.0n | 30.0 | 0.001 | 0.00 | 0.00 | 0.00 | CB | 5.2 | 5.8 | 0.00 | 29.572 | 7.943 | - | 0.245 | -3.780 | -4.186 | -3.307 | 0.000 | 0.000 | 0.000 |
| B30f0.3n | 30.0 | 0.001 | 278.96 | 0.30 | 71.26 | CB | 7.3 | 7.7 | 0.00 | 27.404 | 14.255 | - | 0.606 | -5.085 | -3.226 | -4.005 | 1.311 | 1.248 | 5.907 |
| B30f0.4h | 30.0 | 0.001 | 368.54 | 0.40 | 92.56 | OB | 7.8 | 8.1 | 1.59 | 20.076 | 16.878 | - | 0.309 | -0.386 | -2.966 | -0.556 | 3.371 | 8.802 | 52.199 |
| B30f0.5h | 30.0 | 0.001 | 454.80 | 0.50 | 111.55 | NE | 7.9 | 8.3 | 1.78 | 16.326 | 13.472 | - | 0.080 | -0.423 | -6.812 | -0.267 | 1.110 | 4.488 | 21.686 |
| B30f0.8h | 30.0 | 0.001 | 696.86 | 0.80 | 158.29 | OB | 8.4 | 8.8 | 1.85 | 15.107 | 12.297 | - | 0.092 | -0.395 | -7.227 | -0.298 | 0.919 | 3.891 | 18.826 |
| B40f0.0n | 40.0 | 0.001 | 0.00 | 0.00 | 0.00 | CE | 4.3 | 4.7 | 0.00 | 38.857 | 12.412 | - | 0.261 | -3.837 | -3.951 | -3.334 | 0.000 | 0.000 | 0.000 |
| B40f0.1n | 40.0 | 0.001 | 99.99 | 0.10 | 41.40 | CE | 4.3 | 4.7 | 0.00 | 38.664 | 12.100 | - | 0.258 | -3.979 | -3.798 | -3.342 | 5.668 | 1.192 | 4.963 |
| B40f0.2n | 40.0 | 0.001 | 199.25 | 0.20 | 81.95 | CE | 5.3 | 5.7 | 0.00 | 35.533 | 17.044 | - | 0.543 | -4.735 | -3.267 | -3.843 | 2.560 | 1.280 | 6.787 |
| B40f0.3h | 40.0 | 0.001 | 296.98 | 0.30 | 120.75 | OB | 5.8 | 6.2 | 1.31 | 19.894 | 16.952 | 0.068 | 0.045 | -0.552 | - | -0.173 | 0.736 | 2.374 | 12.977 |
| B40f0.5h | 40.0 | 0.001 | 483.78 | 0.50 | 188.68 | OB | 6.0 | 6.4 | 1.48 | 16.318 | 13.473 | 0.073 | 0.069 | -0.462 | - | -0.234 | 0.389 | 1.817 | 8.028 |
| B40f0.8h | 40.0 | 0.001 | 740.34 | 0.80 | 267.03 | OE | 6.3 | 6.7 | 1.58 | 15.786 | 12.950 | 0.074 | 0.074 | -0.444 | - | -0.249 | 0.404 | 1.963 | 8.466 |
| B60f0.0n | 60.0 | 0.001 | 0.00 | 0.00 | 0.00 | CB | 3.3 | 3.7 | 0.00 | 52.457 | 21.798 | - | 0.373 | -4.055 | -3.494 | -3.522 | 0.000 | 0.000 | 0.000 |
| B60f0.1n | 60.0 | 0.001 | 109.35 | 0.10 | 85.95 | CB | 3.5 | 3.9 | 0.00 | 49.920 | 22.238 | - | 0.455 | -4.241 | -3.382 | -3.637 | 5.326 | 1.278 | 8.295 |
| B60f0.2h | 60.0 | 0.001 | 217.85 | 0.20 | 170.05 | CB | 4.3 | 4.6 | 0.63 | 49.379 | 42.603 | 6.491 | 0.887 | -5.077 | -3.173 | -4.973 | 2.575 | 1.777 | 19.306 |
| B60f0.3h | 60.0 | 0.001 | 323.83 | 0.30 | 248.54 | OE | 4.2 | 4.6 | 1.20 | 19.584 | 18.867 | 0.046 | 0.049 | -0.567 | - | -0.169 | 0.242 | 0.954 | 5.698 |

**Table 7.** Same as in Table 4, but for $Z = 0.00001$.

| No. | $M_{\rm init}$ [$M_\odot$] | $Z_{\rm init}$ | $v_{\rm init}$ [kms$^{-1}$] | $v_{\rm init}/v_{\rm K}$ | $J_{\rm init}$ $10^{51}$ [ergs$^{-1}$] | End | $t_{\rm MS}$ $10^6$ [yr] | $t_{\rm f}$ $10^6$ [yr] | $t_{\rm WR}$ $10^6$ [yr] | $M_{\rm f}$ [$M_\odot$] | $M_{\rm CO}$ [$M_\odot$] | $\Delta M_{\rm He}$ [$M_\odot$] | $Y_{\rm s}$ | $X_{\rm C}$ | $X_{\rm N}$ | $X_{\rm O}$ | $J_{\rm f}$ $10^{51}$ [erg/s] | $< j >_{3M_\odot}$ $10^{15}$ [cm$^2$/s] | $< j >_{\rm CO}$ $10^{15}$ [cm$^2$/s] |
|---|---|---|---|---|---|---|---|---|---|---|---|---|---|---|---|---|---|---|---|
| C12f0.0n | 12.0 | $10^{-5}$ | 0.00 | 0.00 | 0.00 | CB | 13.5 | 15.4 | 0.00 | 11.999 | 1.737 | - | 0.240 | -5.798 | -6.171 | -5.303 | 0.000 | 0.000 | 0.000 |
| C12f0.5n | 12.0 | $10^{-5}$ | 439.72 | 0.50 | 18.34 | CC | 17.2 | 18.7 | 0.00 | 11.949 | 2.105 | - | 0.296 | -7.325 | -5.235 | -6.011 | 16.134 | 1.496 | 1.072 |
| C12f0.6h | 12.0 | $10^{-5}$ | 519.72 | 0.60 | 21.06 | OB | 25.6 | 26.2 | 3.09 | 10.026 | 7.826 | 1.532 | 0.909 | -1.127 | -2.596 | -1.891 | 2.029 | 5.754 | 27.631 |
| C12f0.7h | 12.0 | $10^{-5}$ | 598.06 | 0.70 | 23.63 | NE | 26.0 | 26.7 | 3.24 | 9.908 | 7.440 | 1.046 | 0.990 | -2.376 | -2.547 | -2.601 | 1.890 | 4.743 | 25.756 |
| C12f0.8h | 12.0 | $10^{-5}$ | 675.74 | 0.80 | 26.18 | CE | 26.6 | 27.2 | 3.41 | 9.880 | 7.625 | 1.086 | 0.658 | -0.638 | -2.775 | -0.960 | 2.029 | 8.204 | 30.420 |
| C16f0.0n | 16.0 | $10^{-5}$ | 0.00 | 0.00 | 0.00 | CE | 9.2 | 10.4 | 0.00 | 15.998 | 2.684 | - | 0.240 | -5.759 | -6.288 | -5.302 | 0.000 | 0.000 | 0.000 |
| C16f0.3n | 16.0 | $10^{-5}$ | 284.08 | 0.30 | 20.08 | CE | 9.5 | 10.7 | 0.00 | 15.997 | 2.733 | - | 0.248 | -7.405 | -5.363 | -5.600 | 20.022 | 1.225 | 1.018 |
| C16f0.4n | 16.0 | $10^{-5}$ | 375.56 | 0.40 | 26.12 | CE | 10.3 | 11.4 | 0.00 | 15.995 | 2.982 | - | 0.272 | -7.367 | -5.286 | -5.783 | 25.958 | 1.239 | 1.217 |
| C16f0.5h | 16.0 | $10^{-5}$ | 463.88 | 0.50 | 31.55 | NE | 15.8 | 16.3 | 2.03 | 13.930 | 11.247 | 2.143 | 0.945 | -1.469 | -1.847 | -2.426 | 4.379 | 7.974 | 59.957 |
| C16f0.6h | 16.0 | $10^{-5}$ | 548.16 | 0.60 | 36.22 | OB | 16.1 | 16.6 | 2.32 | 13.335 | 10.285 | 0.606 | 0.967 | -1.849 | -2.316 | -1.852 | 4.549 | 9.968 | 63.371 |
| C16f0.8h | 16.0 | $10^{-5}$ | 712.60 | 0.80 | 45.04 | NE | 16.7 | 17.3 | 2.48 | 13.398 | 10.298 | 0.527 | 0.365 | -0.447 | -2.895 | -0.563 | 5.221 | 14.359 | 76.137 |
| C20f0.0n | 20.0 | $10^{-5}$ | 0.00 | 0.00 | 0.00 | CE | 7.2 | 8.0 | 0.00 | 19.996 | 4.012 | - | 0.240 | -5.759 | -6.288 | -5.302 | 0.000 | 0.000 | 0.000 |
| C20f0.3n | 20.0 | $10^{-5}$ | 296.90 | 0.30 | 30.48 | CE | 8.1 | 8.9 | 0.00 | 19.993 | 4.410 | - | 0.294 | -7.317 | -5.299 | -5.725 | 30.230 | 1.045 | 1.962 |
| C20f0.4n | 20.0 | $10^{-5}$ | 392.44 | 0.40 | 39.64 | CE | 8.7 | 9.4 | 0.00 | 19.857 | 4.812 | - | 0.345 | -7.276 | -5.227 | -6.030 | 31.780 | 1.052 | 2.168 |
| C20f0.45h | 20.0 | $10^{-5}$ | 439.04 | 0.45 | 43.88 | OB | 11.3 | 11.7 | 1.68 | 17.217 | 13.582 | 0.937 | 0.983 | -2.471 | -1.997 | -2.492 | 6.947 | 17.473 | 97.682 |
| C20f0.6h | 20.0 | $10^{-5}$ | 572.56 | 0.60 | 54.92 | OB | 11.5 | 12.0 | 1.91 | 16.539 | 13.008 | 0.645 | 0.412 | -0.522 | -3.057 | -0.545 | 7.407 | 17.566 | 92.902 |
| C20f0.8h | 20.0 | $10^{-5}$ | 744.13 | 0.80 | 68.29 | OB | 12.0 | 12.5 | 2.08 | 16.038 | 12.421 | 0.830 | 0.981 | -2.049 | -2.369 | -2.298 | 6.141 | 14.472 | 81.219 |
| C25f0.0n | 25.0 | $10^{-5}$ | 0.00 | 0.00 | 0.00 | CE | 5.7 | 6.4 | 0.00 | 24.993 | 5.835 | - | 0.240 | -5.759 | -6.288 | -5.302 | 0.000 | 0.000 | 0.000 |
| C25f0.3n | 25.0 | $10^{-5}$ | 310.86 | 0.30 | 46.08 | CE | 6.8 | 7.4 | 0.00 | 24.893 | 7.291 | - | 0.350 | -7.295 | -5.269 | -5.816 | 38.595 | 1.050 | 3.494 |
| C25f0.35n | 25.0 | $10^{-5}$ | 361.22 | 0.35 | 53.15 | NB | 7.6 | 8.1 | 0.00 | 24.457 | 8.656 | - | 0.493 | -7.203 | -5.219 | -6.086 | 27.541 | 1.076 | 4.184 |
| C25f0.4h | 25.0 | $10^{-5}$ | 410.84 | 0.40 | 59.92 | OB | 8.3 | 8.7 | 1.44 | 21.631 | 17.260 | 0.858 | 0.970 | -2.012 | -2.067 | -1.947 | 10.422 | 18.732 | 110.323 |
| C25f0.5h | 25.0 | $10^{-5}$ | 507.23 | 0.50 | 72.32 | OE | 8.4 | 8.8 | 1.59 | 20.966 | 16.916 | 0.596 | 0.516 | -0.635 | -2.639 | -0.604 | 10.882 | 13.592 | 120.435 |
| C25f0.8h | 25.0 | $10^{-5}$ | 778.33 | 0.80 | 103.13 | OE | 8.9 | 9.4 | 1.77 | 19.838 | 15.646 | 0.551 | 0.348 | -0.464 | -3.017 | -0.515 | 10.238 | 19.668 | 106.607 |
| C40f0.0n | 40.0 | $10^{-5}$ | 0.00 | 0.00 | 0.00 | NE | 3.8 | 4.3 | 0.00 | 39.979 | 12.049 | - | 0.240 | -5.759 | -6.288 | -5.302 | 0.000 | 0.000 | 0.000 |
| C40f0.2n | 40.0 | $10^{-5}$ | 230.76 | 0.20 | 73.77 | NE | 4.4 | 4.8 | 0.00 | 39.729 | 14.540 | - | 0.395 | -7.223 | -5.301 | -5.697 | 43.616 | 1.410 | 6.674 |
| C40f0.25n | 40.0 | $10^{-5}$ | 287.66 | 0.25 | 91.52 | YE | 4.9 | 5.3 | 0.00 | 38.984 | 19.078 | - | 0.600 | -7.143 | -5.219 | -6.070 | 28.199 | 2.125 | 11.487 |
| C40f0.3h | 40.0 | $10^{-5}$ | 344.00 | 0.30 | 108.77 | OB | 4.9 | 5.3 | 1.00 | 35.303 | 31.720 | 2.345 | 0.831 | -0.919 | -2.145 | -1.422 | 21.824 | 19.035 | 213.702 |
| C40f0.5h | 40.0 | $10^{-5}$ | 560.67 | 0.50 | 170.32 | OB | 5.0 | 5.3 | 1.21 | 32.988 | 27.974 | 1.030 | 0.620 | -0.811 | -2.509 | -0.656 | 21.465 | 21.249 | 199.338 |
| C40f0.8h | 40.0 | $10^{-5}$ | 858.79 | 0.80 | 242.16 | OB | 5.3 | 5.6 | 1.31 | 31.436 | 27.596 | 1.454 | 0.975 | -1.958 | -2.394 | -2.003 | 18.502 | 19.755 | 192.569 |
| C60f0.0n | 60.0 | $10^{-5}$ | 0.00 | 0.00 | 0.00 | OB | 2.9 | 3.3 | 0.00 | 59.947 | 21.219 | - | 0.240 | -5.759 | -6.288 | -5.302 | 0.000 | 0.000 | 0.000 |
| C60f0.1n | 60.0 | $10^{-5}$ | 126.75 | 0.10 | 77.17 | CE | 3.0 | 3.4 | 0.00 | 59.923 | 21.810 | - | 0.284 | -6.497 | -5.490 | -5.431 | 72.013 | 1.401 | 9.201 |
| C60f0.2n | 60.0 | $10^{-5}$ | 252.54 | 0.20 | 152.75 | CB | 3.5 | 3.9 | 0.18 | 58.598 | 35.098 | - | 0.707 | -7.091 | -5.190 | -6.396 | 29.648 | 2.438 | 21.590 |
| C60f0.3h | 60.0 | $10^{-5}$ | 376.31 | 0.30 | 225.02 | OB | 3.5 | 3.8 | 0.91 | 51.519 | 45.748 | 1.283 | 0.411 | -0.527 | -2.555 | -0.543 | 40.831 | 19.980 | 282.234 |
| C60f0.5h | 60.0 | $10^{-5}$ | 612.43 | 0.50 | 351.16 | OB | 3.5 | 3.9 | 1.06 | 48.444 | 43.019 | 1.154 | 0.376 | -0.482 | -2.993 | -0.535 | 37.761 | 20.672 | 277.912 |
| C60f0.8h | 60.0 | $10^{-5}$ | 935.80 | 0.80 | 496.53 | NE | 3.7 | 4.0 | 1.10 | 46.143 | 40.721 | 1.104 | 0.370 | -0.481 | -3.018 | -0.528 | 35.411 | 20.592 | 269.144 |

**Table 8.** Evolution of central temperature ($T_c$) and density ($\rho_c$), mean toroidal and radial magnetic fields ($< B >_\phi$ & $< B >_r$) and mean specific angular momentum ($< j >$) of the innermost region (1.4 $M_\odot$ and 3.0 $M_\odot$) in selected sequences (S25f0.6h, A25f0.5h, B25f0.5h & C25f0.5h)

| Evolution Stage | $T_c$ [$10^8$K] | $\log \rho_c$ [$cm^3 g^{-1}$] | $< j >_{1.4M_\odot}$ [$10^{15}$ $cm^2 s^{-1}$] | $< B_\phi >_{1.4M_\odot}$ [G] | $< B_r >_{1.4M_\odot}$ [G] | $< j >_{3.0M_\odot}$ [$10^{15}$ $cm^2 s^{-1}$] | $< B_\phi >_{3.0M_\odot}$ [G] | $< B_r >_{3.0M_\odot}$ [G] |
|---|---|---|---|---|---|---|---|---|
| | | | | – S25f0.6h – | | | | |
| He exhaustion | 5.4 | 3.92 | 2.25 | $4.44 \times 10^5$ | $2.03 \times 10^2$ | 4.05 | $6.05 \times 10^5$ | $2.25 \times 10^2$ |
| C exhaustion | 14.9 | 6.12 | 1.76 | $8.15 \times 10^7$ | $1.92 \times 10^4$ | 3.27 | $7.53 \times 10^7$ | $1.56 \times 10^4$ |
| O exhaustion | 26.2 | 7.16 | 1.77 | $4.27 \times 10^8$ | $8.78 \times 10^4$ | 2.95 | $3.67 \times 10^8$ | $6.80 \times 10^4$ |
| | | | | – A25f0.5h – | | | | |
| He exhaustion | 5.3 | 3.85 | 7.69 | $7.10 \times 10^5$ | $4.49 \times 10^2$ | 13.6 | $9.60 \times 10^5$ | $4.92 \times 10^2$ |
| C exhaustion | 14.6 | 5.95 | 5.21 | $1.84 \times 10^8$ | $5.66 \times 10^4$ | 9.99 | $2.04 \times 10^8$ | $5.52 \times 10^4$ |
| O exhaustion | 27.0 | 7.02 | 5.13 | $1.17 \times 10^9$ | $4.21 \times 10^5$ | 9.46 | $1.46 \times 10^9$ | $5.91 \times 10^5$ |
| | | | | – B25f0.5h – | | | | |
| He exhaustion | 4.2 | 3.51 | 11.5 | $4.17 \times 10^5$ | $3.21 \times 10^2$ | 20.3 | $5.78 \times 10^5$ | $3.59 \times 10^2$ |
| C exhaustion | 14.4 | 5.89 | 6.94 | $2.23 \times 10^8$ | $7.41 \times 10^4$ | 13.3 | $2.47 \times 10^8$ | $7.19 \times 10^4$ |
| O exhaustion | 27.4 | 7.06 | 6.55 | $2.21 \times 10^9$ | $1.35 \times 10^6$ | 12.1 | $2.60 \times 10^9$ | $1.75 \times 10^6$ |
| | | | | – C25f0.5h – | | | | |
| He exhaustion | 5.14 | 3.76 | 22.2 | $1.51 \times 10^6$ | $1.26 \times 10^3$ | 39.0 | $2.08 \times 10^6$ | $1.40 \times 10^3$ |
| C exhaustion | 14.2 | 5.77 | 11.5 | $2.95 \times 10^8$ | $1.11 \times 10^5$ | 22.0 | $3.39 \times 10^8$ | $1.14 \times 10^5$ |
| O exhaustion | 30.9 | 7.03 | 10.0 | $4.69 \times 10^9$ | $5.20 \times 10^6$ | 19.2 | $6.52 \times 10^9$ | $8.49 \times 10^6$ |